# Quantum Chaos Control by Complex Trajectories

*Ciann-Dong Yang [a], Yen-Jiun Chen [a], Yun-Yan Lee [a,\*]*

*"Department of Aeronautics and Astronautics, National Cheng Kung University, Taiwan"*

ABSTRACT

In recent years, analysis and control of quantum chaos are increasingly important, but the lack of the concept of trajectory makes it impossible to analyze quantum chaos by the methods used in classical chaos. This research aims to connect Newton's world to the quantum world by the complex mechanics so that quantum chaos can be analyzed and controlled by the complex-extended Newtonian mechanics. Through the bridge of complex mechanics, in this article, we model quantum motions for 2D charged anisotropic harmonic oscillator by complex-valued dynamic equations, based on which quantum chaos can be analyzed by using well-known methods used in classical chaos. With the established quantum dynamic model, we then apply the sliding-mode control method to control the chaotic quantum behavior of the considered quantum system. The simulation results show that chaotic motions can be changed into periodic motions by the proposed chaos control and meanwhile, chaos synchronization can be achieved in the presence of variations of initial conditions. Several signatures of chaos are introduced here to justify the chaos of the periodicity process under the sliding-mode control law.

## 1. Introduction

Quantum control is a new popular research area since the 21st century. The main progress of quantum control is to control the microscopic state of atoms and molecules. Although experiments related to quantum control have been a hot topic in scientific journals in recent years, the idea of quantum control originated very early. Around the 1980s, the research on quantum systems progressed from theoretical analysis to practical applications. In 1983, G.M. Huang and T.J. Tarn brought up the research on the controllability of quantum systems [1], analyzing the conditions under which quantum systems can be controlled under limited space dimensions. N. Bloembergen also proposed the idea of quantum control the following year, but due to the immaturity of laser control technology at that time, his thought did not get people's attention. In 1990, H. Rabitz proposed a specific quantum control idea and used it in the Hamiltonian of the carbon-hydrogen bond, and then in 1992 H. Rabitz continued to put forward the idea of learning control in the journal Physical Review Letters [2], and mentioned in the article that people can observe the signals generated by the interaction of light and molecules, such as fluorescence and dissociation, etc. Then quickly analyze this signal and compare it with the target state which people expect.

The idea that feedback to the laser and change the parameters to make the molecule reach the desired state is called "learning control".

This kind of close-loop control research gradually got rid of the previous idea of open-loop control of quantum systems, coupled with the development of laser control technology, molecular quantum control experiments became more and more mature, so during the 1990s, there were continuous problems. Few relevant experimental results, e.g., in 1998, A. Assion et al. used voltage to adjust the spatial distribution of liquid crystal refractive index to reshape pulses, and used computers to perform learning control and successfully proved that pulses can be used to control the process and products of chemical changes [3]. In 2000, J. Kunde demonstrated the use of pulse shaping and feedback control to adjust the nonlinear characteristics of semiconductors [4].

Like the development of classical control theory, quantum control will also shift from learning control to feedback control. The purpose of learning control is whether to acquire a certain product we want, while feedback control hopes to remain in the desired state forever. However, the challenge for the development of quantum feedback control lies in the unique Heisenberg uncertainty principle of quantum systems, which is caused by noise in measurement, system modeling, and interaction between signals and controlled systems. Uncertainties and disturbances make the state of the quantum system unable to be accurately measured. At the end of the 20th century and the beginning of the 21st century, due to the breakthrough in measurement theory and technology, studies on the measurement of the state of quantum systems and the combination of classical control theories began one after another.

In 1999, A.C. Doherty proposed a method to estimate the state of the system by analyzing the feedback variables, and compared it with the classical LQG (Linear Quadratic Gaussian) control principle [5], confirming the feasibility of the feedback control of the quantum system. Besides, M. Yanagisawa and H. Kimura proposed in 2003 that quantum systems can use operators to obtain transfer functions [6,7]. The feedback control method of the closed-loop control of the quantum system has gradually developed with the progress of quantum measurement technology, the feedback control process of the closed-loop of the quantum system [8]. As shown

\* *Corresponding author.* Tel.: +886-934-066-169
E-mail address: d187108@gmail.com

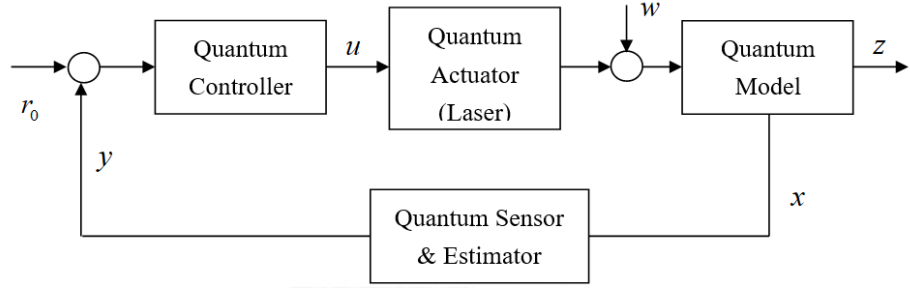

**Figure 1.1 The structure of quantum system feedback control**

When applying feedback control to quantum systems, how to improve the robust performance of the system is the core research directions of quantum control because the robust control can have excellent resistance to the uncertainty and disturbance of the quantum system.

On the other hand, classical chaos control theory has flourished. In 1990, L.M. Pecora and T.M. Carroll proposed a method for synchronizing two identical chaotic systems under different initial conditions [9], chaotic synchronization has played a critical role in nonlinear science in the past few decades. Many scientists are dedicated to the study of chaotic synchronization in different fields as secure communication, optics, and chemical engineering. Therefore, many methods for chaos synchronization have been proposed [10,11,12], such as linear feedback control, adaptive control and Active control, etc. Nonlinear control method. This article uses sliding control to achieve chaotic synchronization. M.S. Tavazoei and M. Haeri use sliding control to achieve chaotic synchronization in 2008. [13] The advantage of this method is that we first perform bang-bang control on the sliding surface. This step makes the problem to be a one-dimensional nonlinear control problem, which is simpler than directly using Lyapunov control to synchronize chaotic phenomenon. [14]. When the state reaches the sliding surface, it becomes a linear control problem. The sliding control is also used in quantum mechanics, C. Yang et.al. uses sliding control to synchronize polarizations and phase displacements, [15]. Afterwards, the discussion of quantum chaos synchronization also directly discussed the evolution of quantum. [16][17] On the other hand, in the quantum sliding control, D. Dong et. al. directly use quantum states to establish sliding surfaces. [18][19] [20] S. Chegini and M. Yarahmadi improve this method. [21] However, this article transforms a quantum control problem to a classical control problem with complex mechanics which we will introduce later, and we use the sliding control to deal with the classical control problem based on complex mechanics.

Classical control theory is not easy to strike a balance between rapid system response and robustness, and the parameters of the controlled body are often changed by external disturbances, making it difficult for the design of traditional controllers to meet all conditions. The sliding mode control theory [22] not only makes the system respond faster but also the designed controller has better robustness, because of its better resistance to uncertainty and noise. Its characteristic is that the control gain will switch between different values, and different gains represent different structures, so it is also called a variable structure system (VSS). Sliding mode control consists of three steps:

(1) Select the sliding $s(\mathbf{x})$ surface function.
(2) The design of the sliding mode controller $u$.
(3) Analysis of robustness and stability.

The sliding surface is a curve or curved surface in the phase plane. It must be a once differentiable and contain a balance point. It represents the final state that the control system will reach. In other words, sliding control is the control that drives the system into the sliding surface. Suppose that there is an n-order phase plane, there is a sliding surface $s(\mathbf{x}) = 0$, and the control process has the following two stages:

(1) No matter where the system dynamics are in the phase plane, there is a tendency to move to the sliding surface under the sliding control strategy. That is, the sliding surface $s(\mathbf{x})$ must satisfy $\dot{s}s < 0$, which is called the proximation condition.
(2) When the system enters the sliding surface, the sliding control strategy ensures that the system remains on the sliding surface and cannot leave. That is, the sliding surface must satisfy $\dot{s}(\mathbf{x}) = 0$, which is called the sliding condition.

When selecting the sliding surface $s(\mathbf{x}) = 0$, the system state space separates into two subspaces $s(x) > 0$ and $s(x) < 0$. When designing a sliding controller, the control input $u$ forces the sliding surface to satisfy the proximation condition and the sliding condition. The corresponding sliding controller has two parts. One part is used in the subspace outside the sliding surface as the switching control $u_N$ so that it satisfies the formula $\dot{s}s < 0$, so the switching logic is

$$u_N = \begin{cases} u_N^+ & s(\mathbf{x}) < 0 \\ u_N^- & s(\mathbf{x}) > 0 \end{cases}$$
(1.1)

The other part is when the system enters the sliding surface, making a balance control $u_{Eq}$ such that the condition $\dot{s}(\mathbf{x}) = 0$ is satisfied. The system can enter the equilibrium point without leaving the sliding surface. Therefore, complete control input can be regarded as switching control $u_N$ and balance control $u_{Eq}$. This is the design rule of the sliding controller. Figure 1.2 is a schematic diagram of a sliding mode control.

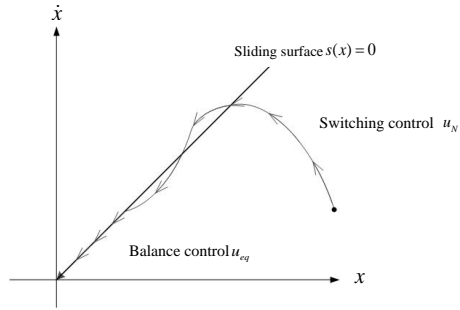

**Figure 1.2 Schematic diagram of the sliding mode control**

When the sliding control law is determined, we can use the Lyapunov stability analysis to confirm the existence of a certain Lyapunov function, so that the system satisfies in the closed-loop situation and achieves system stability.

In quantum mechanics, the motion of particles is usually described by a probability distribution, but there is no equation of motion used to find the trajectory of the particle, so the period or frequency of the particle's motion is usually can only be obtained indirectly by experiment, which makes quantum chaotic behavior unable to be defined and analyzed by traditional methods. To solve this problem, we introduced the concept of complex mechanics [23,24,25,26,27] and combined it with quantum mechanics, dividing position and momentum into real and imaginary parts. Since the real world we live in is in the real three-dimensional space, we can't observe the movement of particles in the complex space. With the bridge constructed by complex mechanics, we can obtain the equation of motion of the particle in the complex space, and further obtain the trajectory of the particle from the equation. From the divergent form of adjacent trajectories, it can present phenomena that are generally invisible in quantum chaos. And the potential energy of qubits introduced by complex mechanics can be used to explain the reason behind the probability distribution of particles.

Complex mechanics is mainly to establish the equivalent relationship between the complex physical quantity $A(\mathbf{q},\mathbf{p})$ in Hamilton mechanics and the corresponding quantum operator in quantum mechanics, where position and momentum $(\mathbf{q},\mathbf{p})$ are defined in the complex space. In this way, quantum mechanics can be re-described by Hamilton's equation in standard complex variables. The redefined Hamilton equation is different from the classical Hamilton $H_c(\mathbf{q},\mathbf{p},t)$. The classical Hamilton-Jacobi (H-J) equation is

$$\frac{\partial S_c}{\partial t} + H_c(\mathbf{q},\mathbf{p},t)\,|_{\mathbf{p}=\nabla S_c} = 0 \tag{1.2}$$

Where $S_c$ is the classical action function. We can regard the classical H-J equation as the short-wavelength limitation of Schrodinger's equation [28]

$$i\hbar\frac{\partial \psi}{\partial t} = -\frac{\hbar^2}{2m}\nabla^2\psi + V\psi \tag{1.3}$$

Where $\psi$ is the wave function, $\hbar$ is Planck's constant, and $V$ is the applied potential. Substitute $\psi = \exp(iS/\hbar)$ and we can get quantum H-J equation

$$\frac{\partial S}{\partial t} + [\frac{1}{2m}(\nabla S)^2 + V] = \frac{i\hbar}{2m}\nabla^2 S \tag{1.4}$$

If Keeping the item on RHS, this item can be ignored in the macro world, but in the atomic level micro world, it can't be ignored. Rewrite Eq. (1.4) as the type of Eq. (1.2):

$$\frac{\partial S}{\partial t} + \left[\frac{1}{2m}(\nabla S)^2 + V - \frac{i\hbar}{2m}\nabla^2 S\right] = \frac{\partial S}{\partial t} + \left[\frac{\mathbf{P}^2}{2m} + V + \frac{\hbar}{2mi}\nabla\cdot\mathbf{p}\right]\Bigg|_{\mathbf{p}=\nabla S} = 0 \tag{1.5}$$

The revised Hamilton is compatible with Schrodinger's equation, so we can obtain the desired quantum Hamilton as

$$H(\psi) = \frac{1}{2m}\mathbf{p}^2 + V(t,\mathbf{q}) + Q(\psi(t,\mathbf{q})) \tag{1.6}$$

The third term of RHS $Q = (\hbar/2mi)\nabla\cdot\mathbf{p}$ is defined as the quantum potential, because $\mathbf{p} = \nabla S$ and $S = (\hbar/i)\,\ln\,\psi$, so $Q$ can be expressed in the form of the wave function:

$$Q(\psi(t,\mathbf{q})) = \frac{\hbar}{2mi}\nabla\cdot\mathbf{p} = \frac{\hbar}{2mi}\nabla^2 S = -\frac{\hbar^2}{2m}\nabla^2\ln\psi(t,\mathbf{q}) \tag{1.7}$$

Quantum potential plays an important role in the microscopic size, and after adding the applied potential, it can completely determine the movement path of the particle. After merging the additional potential and quantum potential, a new total potential can be obtained:

$$V_{Total}(t,\mathbf{q}) = V(t,\mathbf{q}) + Q(\psi(t,\mathbf{q})) \tag{1.8}$$

Therefore, the reason why the traditional energy conservation is inconsistent in quantum mechanics is that the term quantum potential is missing, so that the total energy of the particle cannot be accurately described. When we get the total energy of real particles, we can apply Eq. (1.5) to classical mechanics.

In quantum Hamilton, the wave function provides two important physical quantities, namely canonical momentum, which can be determined by the following formula

$$\frac{d\mathbf{q}}{dt} = \frac{\partial H(\psi)}{\partial \mathbf{p}} = \frac{1}{m}\mathbf{p} \quad,\mathbf{q}\in\quad^3 \tag{1.9a}$$

$$\frac{d\mathbf{p}}{dt} = -\frac{\partial H(\psi)}{\partial \mathbf{q}} = -\frac{\partial}{\partial \mathbf{q}}\left[V(t,\mathbf{q}) - \frac{\hbar^2}{2m}\nabla^2 \ln\psi(t,\mathbf{q})\right] \quad ,\mathbf{p} \in \quad^3$$

(1.10b)

It should be noted that, compared with classical Hamilton, the difference is that the two independent variable solutions of the quantum Hamilton equation are both complex numbers and are uniquely determined by the wave function that dominates the quantum motion.

## 2. Quantum Harmonic oscillator with Complex Mechanics

For a charged particle in the anisotropic harmonic potential $V(r)$ with a uniform magnetic field in the z-direction $\boldsymbol{B} = Be_z$, its classical Hamilton can be expressed as

$$H = \frac{1}{2m}(\mathbf{p} - q\mathbf{A})^2 + V(r) \quad \mathbf{p},\mathbf{r} \in \quad^3$$

(2.1)

Where $m$ is the particle mass, $\mathbf{p}$ and $\mathbf{r}$ respectively represent the momentum and position of the particle. The second term of RHS is a three-dimensional simple harmonic potential

$$V(r) = \frac{m}{2}(\omega_x^2 x^2 + \omega_y^2 y^2 + \omega_z^2 z^2) \quad x,y,z \in$$

(2.2)

For a uniform magnetic field $\boldsymbol{B}$, the vector potential $\boldsymbol{A}$ can be obtained as

$$\mathbf{A} = \frac{-B}{2}(y\vec{i} - x\vec{j})$$

(2.3)

$i$ and $j$ are the unit vectors along the axis-$x$ and axis-$y$ respectively. The above system is defined by classical mechanics. Because quantum effects are ignored, the physical quantities have not been in complex space and the quantum potential has not been considered. After substituting Eq. (2.2) and Eq. (2.3) into Eq. (2.1), the entire Hamilton can be decomposed into the two-dimensional vertical magnetic field $H_{xy}$ and the one-dimensional simple harmonic oscillator $H_z$. That is $H = H_{xy} + H_z$, where

$$H_{xy} = \frac{1}{2m}(p_x^2 + p_y^2) + \frac{\omega_c}{2}(xp_y - yp_x) + \frac{m\omega_c^2}{8}(x^2 + y^2) + \frac{m}{2}(\omega_x^2 x^2 + \omega_y^2 y^2)$$

(2.4)

$$H_z = \frac{1}{2m}p_z^2 + \frac{m}{2}\omega_z^2 z^2$$

(2.5)

Where $\omega_c = -qB/m$ is the cyclotron frequency of the magnetic field. Corresponding to the wave function solution of the time-invariant Schrodinger equation in the three-dimensional quantum system, which can be written as the product of two characteristic functions $\psi_{xy}$ and $\psi_z$ respectively.

$$\hat{H}_z\psi_z = E_z\psi_z$$

(2.6a)

$$\hat{H}_{xy}\psi_{xy} = E_{xy}\psi_{xy}$$

(2.6b)

According to the normalization operator proposed by Dippel [29]

$$U = \exp(i\varsigma xy)\exp(i\xi p_x p_y)$$

(2.7)

It can eliminate the angular momentum $\omega_c(xp_y - yp_x)/2$ in Eq. (2.4), and produce a new Hamiltonian, which contains two independent one-dimensional harmonic oscillators.

$$\hat{H}_{xy} = U^{-1}H_{xy}U = \left(\frac{p_x^2}{2M_1} + \frac{M_1}{2}\omega_1^2 x^2\right) + \left(\frac{p_y^2}{2M_2} + \frac{M_2}{2}\omega_2^2 y^2\right)$$

(2.8)

The following conversion relationship is used

$$\varsigma = \rho_1 m\omega_x \quad \xi = \rho_2/(m\omega_x) \quad \omega_{1,2} = \eta_{1,2}\omega_x \quad M_{1,2} = \mu_{1,2}m \quad \gamma = \omega_y/\omega_x \quad \beta = \omega_c/\omega_x$$

$$\rho_1 = \frac{(\gamma^2 - 1) \pm \sqrt{(1 + \gamma^2 + \beta^2)^2 - 4\gamma^2}}{2\beta} \quad \rho_2 = \frac{\pm\beta}{\sqrt{(1 + \gamma^2 + \beta^2)^2 - 4\gamma^2}}$$

$$\eta_{1,2} = \frac{1}{\sqrt{2}}\left[1 + \gamma^2 + \beta^2 \pm sign(1 - \gamma^2)\sqrt{(1 + \gamma^2 + \beta^2)^2 - 4\gamma^2}\right]^{\frac{1}{2}}$$

$$\mu_{1,2} = \frac{2\sqrt{(1 + \gamma^2 + \beta^2)^2 - 4\gamma^2}}{sign(1 - \gamma^2)(1 - \gamma^2 \pm \beta^2) + \sqrt{(1 + \gamma^2 + \beta^2)^2 - 4\gamma^2}}$$

(2.9)

So the new wave function can be regarded as the product of the respective wave functions of the two simple harmonic oscillators, that is

$$\psi'_{n_1 n_2}(x,y) = \phi_{n_1}(x)\phi_{n_2}(y)$$

(2.10)

$$E_{n_1 n_2} = \left(n_1 + \frac{1}{2}\right)\eta_1\hbar\omega_x + \left(n_2 + \frac{1}{2}\right)\eta_2\hbar\omega_y$$

(2.11)

The wave function solution $\psi_{n_1 n_2}$ in Eq. (2.6b) can be obtained by converting the new wave function $\psi'_{n_1 n_2}$ with the normalization operator $U$.

$$\psi_{n_1 n_2}(x,y) = U\psi'_{n_1 n_2}(x,y)$$

$$= e^{i\rho_1\bar{x}^2 + i\alpha_0\bar{x}\bar{y} + i\omega_2\bar{y}^2}\sum_{k=0}^{n_1}\sum_{l=0}^{n_2}c_{kl}(n_1,n_2)H_{n_1-k}\left(\sqrt{2}(a_1\bar{x} + ib_1\bar{y})\right)H_{n_2-l}\left(\sqrt{2}(a_2\bar{x} - ib_2\bar{y})\right)$$

(2.12)

Where $H_n$ is the $n$ order of Hermite polynomial. Because the eigenvalue of $\hat{H}_{xy}$ and the eigenvalue of $H_{xy}$ are the same, the time-dependent wave function can be expressed as [30]

$$\psi_{n_1 n_2}(\bar{x},\bar{y},\bar{t}) = \psi_{n_1 n_2}(\bar{x},\bar{y})e^{-i\bar{E}_{n_1 n_2}\bar{t}}$$

(2.13)

Define the dimensionless variable

$$\bar{t} = \omega_x t \quad \bar{x} = \sqrt{m\omega_x/\hbar}\,x \quad \bar{y} = \sqrt{m\omega_x/\hbar}\,y \quad \bar{E}_{n_1 n_2} = E_{n_1 n_2}/(\hbar\omega_x)$$

And define the following parameters an

$$a_1 = \frac{\sqrt{\mu_1 \eta_1}}{\rho_2^2 \mu_1 \eta_1 \mu_2 \eta_2 + 1} \qquad b_1 = \frac{\sqrt{\mu_1 \eta_1} \rho_2 \mu_2 \eta_2}{\rho_2^2 \mu_1 \eta_1 \mu_2 \eta_2 + 1} \qquad a_2 = \frac{\sqrt{\mu_2 \eta_2} \rho_2 \mu_1 \eta_1}{\rho_2^2 \mu_1 \eta_1 \mu_2 \eta_2 + 1} \qquad b_2 = \frac{\sqrt{\mu_2 \eta_2}}{\rho_2^2 \mu_1 \eta_1 \mu_2 \eta_2 + 1}$$

$$a_3 = \frac{-\mu_1 \eta_1}{2(\rho_2^2 \mu_1 \eta_1 \mu_2 \eta_2 + 1)} \qquad a_4 = \rho_1 - \frac{\rho_2 \mu_1 \eta_1 \mu_2 \eta_2}{\rho_2^2 \mu_1 \eta_1 \mu_2 \eta_2 + 1} \qquad a_5 = \frac{-\mu_2 \eta_2}{2(\rho_2^2 \mu_1 \eta_1 \mu_2 \eta_2 + 1)}$$

$$c_{kl}(n_1, n_2) = \begin{cases} 0, \ for \ k+l \ odd \\[2mm] 2^{2/4} \binom{n_1}{k} \binom{n_2}{l} \left(\frac{k+l+1}{2}\right) D^l G^l (2D^2 - 1)^{(k-l)/2} \\[2mm] \times {}_2F_1\left(-\frac{l}{2}, \frac{1-l}{2}; \frac{1-k-l}{2}; \frac{1}{2D^2} + \frac{1}{2G^2} - \frac{1}{4D^2 G^2}\right), otherwise \end{cases} \tag{2.14}$$

Defined $2F_1(\cdot)$ as a hypergeometric function [31], and the parameters $D, G$ are determined by the following formula

$$D = \sqrt{\frac{2\rho_2^2 \mu_1 \eta_1 \mu_2 \eta_2}{\rho_2^2 \mu_1 \eta_1 \mu_2 \eta_2 + 1}} \qquad G = -sign(1-\gamma^2)\sqrt{\frac{2}{\rho_2^2 \mu_1 \eta_1 \mu_2 \eta_2 + 1}}$$

Then, applying the method of the complex mechanic, we find that the standard momentum $p_x, p_y$ can be determined

$$p_x = \frac{\partial S}{\partial x} = -i\hbar \frac{\partial \ln \psi_{n_1 n_2}}{\partial x} \qquad p_y = \frac{\partial S}{\partial y} = -i\hbar \frac{\partial \ln \psi_{n_1 n_2}}{\partial y} \tag{2.15}$$

Where $x, y, p_x, p_y \in \mathbb{C}$ and S is a complex action function with the following relationship

$$S = \frac{\hbar}{i} \ln \psi \tag{2.16}$$

The wave function $\psi$ is a linear combination of multiple eigenstates

$$\psi = C_0 \psi_{00} + C_1 \psi_{10} + C_2 \psi_{01} + \dots \tag{2.17}$$

The dimensionless particle dynamic equation in the characteristic quantum state $\psi$ can be expressed as

$$\frac{d\bar{x}}{d\bar{t}} = \sqrt{\frac{m}{\omega_x \hbar}} \frac{dx}{dt} = \sqrt{\frac{1}{m\omega_x \hbar}} p_x = -i \frac{\partial \ln \psi}{\partial \bar{x}}, \quad \bar{x} \in$$

$$\frac{d\bar{y}}{d\bar{t}} = \sqrt{\frac{m}{\omega_x \hbar}} \frac{dy}{dt} = \sqrt{\frac{1}{m\omega_x \hbar}} p_y = -i \frac{\partial \ln \psi}{\partial \bar{y}}, \quad \bar{y} \in \tag{2.18}$$

After numerical integration, the two complex equations in Eq. (2.18) can be divided into real and imaginary parts and obtain four time-invariant quantum trajectories in the characteristic state.

Now consider that when the particle is in the ground state, substituting $n_1 = 0, n_2 = 0$ into Eq. (2.10), the wave function of the ground state can be obtained as

$$\psi_{00}(\bar{x}, \bar{y}) = c_{00}(0,0) \exp(a_3(\gamma, \beta) \bar{x}^2 + i a_4(\gamma, \beta) \bar{x}\bar{y} + a_5(\gamma, \beta) \bar{y}^2) \tag{2.19}$$

The coefficients $a_3, a_4, a_5$ can be from Eq. (2.9) (2.14) that are dominated by two parameters, one of which is the ratio of simple harmonic frequency $\gamma$ and the other contains the intensity of the magnetic field $\beta$. With the wave function, the dynamic equation of the quantum trajectory can be determined according to Eq. (2.15)

$$\frac{d\bar{x}}{d\bar{t}} = -i(2a_3(\gamma, \beta) \bar{x} + i a_4(\gamma, \beta) \bar{y}),$$

$$\frac{d\bar{y}}{d\bar{t}} = -i(i a_4(\gamma, \beta) \bar{x} + 2a_5(\gamma, \beta) \bar{y}) \tag{2.20}$$

According to the above formula, we can solve the solution of $\bar{x}$ and $\bar{y}$, and the result is a linear combination of two exponential functions. Similarly, the wave function ($n_1 = 1, n_2 = 0$) of the particle in the first excited state can be obtained as

$$\psi_{10}(\bar{x}, \bar{y}) = 2\sqrt{2} c_{00}(1,0)(a_1(\gamma, \beta) \bar{x} + i b_1(\gamma, \beta) \bar{y})$$

$$\times \exp(a_3(\gamma, \beta) \bar{x}^2 - i a_4(\gamma, \beta) \bar{x}\bar{y} - a_5(\gamma, \beta) \bar{y}^2) \tag{2.21}$$

And get the quantum Hamilton equation for the first excited state

$$\frac{d\bar{x}}{d\bar{t}} = \frac{-i(2a_1 a_3 \bar{x}^2 + 2i a_3 b_1 \bar{x}\bar{y} + i a_1 a_4 \bar{x}\bar{y} - a_4 b_1 \bar{y}^2 + a_1)}{(a_1 \bar{x} + i b_1 \bar{y})}$$

$$\frac{d\bar{y}}{d\bar{t}} = \frac{-i(a_1 a_4 \bar{x}^2 - a_4 b_1 \bar{x}\bar{y} + 2a_1 a_5 \bar{x}\bar{y} + 2i a_5 b_1 \bar{y}^2 + i b_1)}{(a_1 \bar{x} + i b_1 \bar{y})} \tag{2.22}$$

To sum up the above, when we find the wave function of a certain characteristic state under the harmonic potential, the dynamic trajectory equation of the quantum can be determined by Eq. (2.15), and the particles projected on the real two-dimensional plane can also be observed.

When the particle energy level is in the ground state, from the dynamic Eq. (2.20), we assume the parameters $\beta = 0.8, \gamma = 0.4$, and according to Eq. (2.9) and Eq. (2.14), the following parameters can be determined: $a_1 = 0.9868, a_3 = -0.5759, a_4 = 0.1714, a_5 = -0.2304, b_1 = 0.2668$. Let the particles start from $(\bar{x}, \bar{y}) = (1,1)$, which also means that they are moving at the starting point $(x_R, x_I, y_R, y_I) = (1,0,1,0)$ of the two-dimensional complex plane. After 1000 seconds of dynamic simulation, the trajectory is projected on the real number plane $\bar{x}\bar{y}$, and the result is shown in Figure 2.1 below. The figure shows a regular quasi-period movement, and no chaotic phenomenon is revealed.

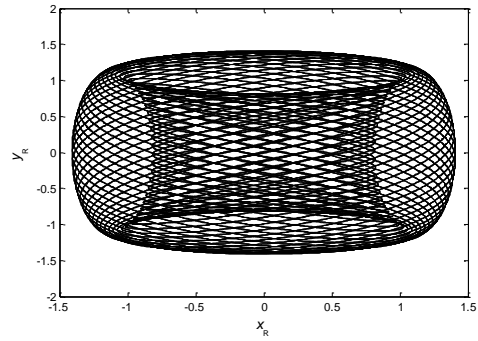

**Figure 2.1 A quasi-periodical trajectory simulated for 1000 seconds in the ground state.**

Also assume the parameters $\beta = 0.8, \gamma = 0.4$, if the energy level is increased to the first excited state $\psi_{10}$, the starting position $(\bar{x}, \bar{y}) = (0.4, 0.8)$ is taken, and the trajectory after 3000 seconds of simulation is shown in Figure 2.2.

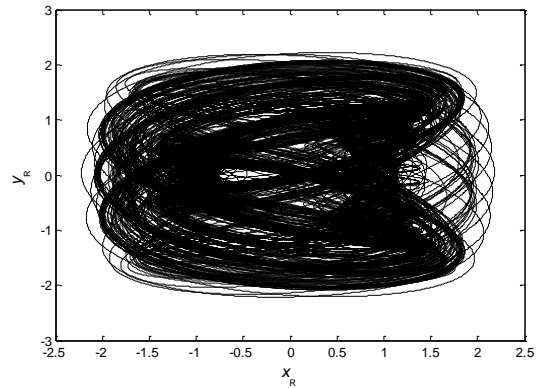

**Figure 2.2 Simulated chaotic trajectory for 3000 seconds in the first excited state**

The above figure shows that in the two-dimensional phase space, the eigenstate has obvious chaotic phenomena.

The classical chaotic phenomenon generally discussed is to perturb the initial position of the trajectory $x_R^0 \rightarrow x_R^0 + \delta x_R^0$, test its sensitivity, and observe the degree of divergence of the trajectory. But if the real part is disturbed $\delta x_R^0 = 0$, the discussion of classical chaos is meaningless, because for classical dynamic systems, $\delta x_R(0) = \delta x_R^0 \equiv 0$ that is, $\delta x_R(t) = \delta x_R(0) e^{\lambda t} \equiv 0, \forall t \geq 0$ the degree of divergence of the representative trajectory is 0, and chaos is impossible. However, for quantum systems, chaotic features may still be discovered without disturbances in the real part. This means that the trajectory may diverge without any disturbance in the initial position. This multipath phenomenon in which multiple trajectories are generated from the same initial position is called strong chaos.

From the perspective of the internal mechanism of multipath behavior, we can infer that strong chaos is mainly caused by the interaction between internal imaginary dynamics and external real dynamics, resulting in an infinite number of trajectories. As shown in Figure 2.3

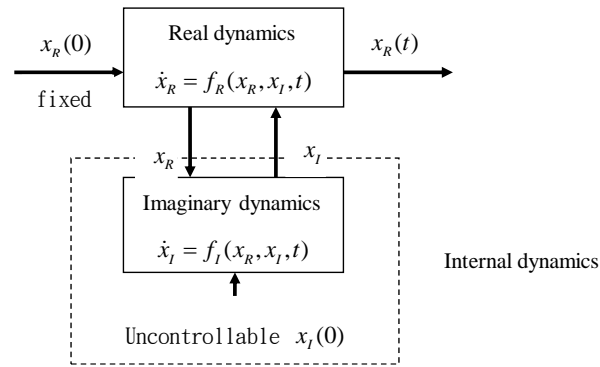

**Figure 2.3 Internal mechanism of the multipath phenomenon**

That is, the initial condition of fixing the real part $x_R(0)$, any imaginary part position $x_I(0)$ can produce a specific trajectory $x_R(t)$. This also implies that the observation space is limited to real numbers. Since the unobservable dynamics of the imaginary part $x_I(0)$ maybe very different, the trajectories start from the same point $x_R(0)$ in the real number phase space, but the differences $x_I(0)$ cause $x_R(t), t > 0$, big differences. Return to the dynamic equation in (2.20), and take the starting positions $(\bar{x}_R, \bar{x}_I, \bar{y}_R, \bar{y}_I)$ as (1,0,0,0) and (1,0.5,0,0.5) respectively, the trajectory drawn on the and axes in Figure 2.4 seems to start from the same point but presents an irrelevant divergence phenomenon, which is mainly affected by the unobserved disturbance of the imaginary part. Figure 2.5 and Figure 2.6 respectively start from the plane trajectories of the same real part and different imaginary parts. It can be understood that under the action of strong chaos, there will be different trajectory patterns.

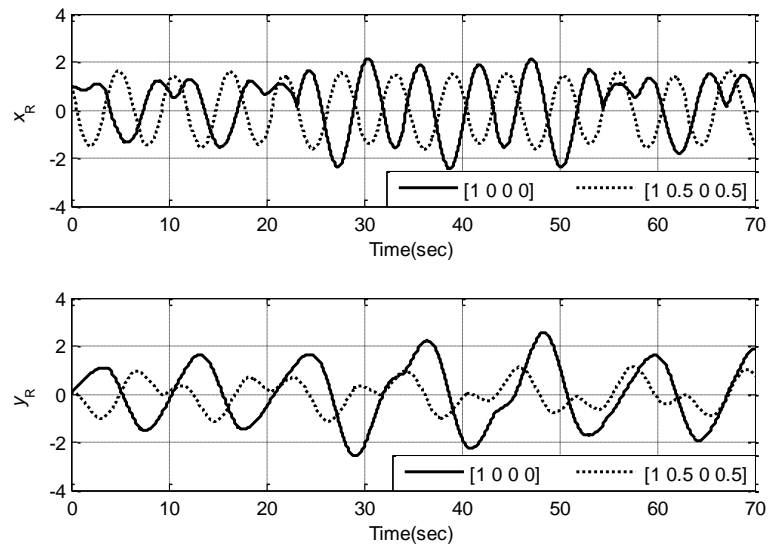

**Figure 2.4 Trajectory divergence caused by the imaginary initial position disturbance**

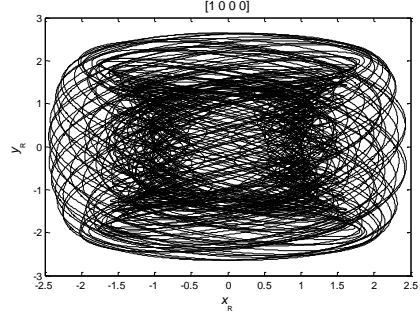

**Figure 2.5 The chaotic plane trajectory from the initial position [1 0 0 0]**

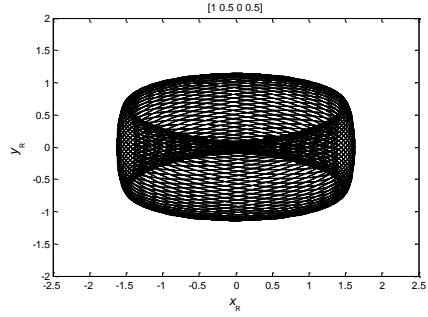

**Figure 2.6 The trajectory of the quasi-period plane starting from the initial position [1 0.5 0 0.5]**

The classical system itself does not dissipate and the physical quantity is conserved. We call it a conserved system, which is equivalent to the Hamilton system. Take the quantum system constructed under quantum Hamilton, the total energy is also conserved. The total energy does not change with time. It is worth thinking about the origin of chaos. According to classical mechanics, the force experienced by particles is governed by an external potential $V$

$$\mathbf{F} = -\nabla V \tag{2.23}$$

When the potential plane distribution is more uniform, the change of the force acting on the particle will be smoother; when the relative potential plane presents a steep undulating distribution, the force will change drastically. The difference from the above is that

In the quantum system, we add additional quantum potential $Q$ to form a total potential $V_{Total}$ to replace the original classical potential $V_{class}$. The quantum potential $Q$ can be determined as

$$Q(\psi(x,y,t)) = -\frac{\hbar}{2m}[\frac{\partial^2 \ln \psi(x,y,t)}{\partial x^2} + \frac{\partial^2 \ln \psi(x,y,t)}{\partial y^2}] \tag{2.24}$$

After dimensionless, it can be rewritten as

$$\bar{Q} = \frac{Q}{\hbar \omega_x} = -\frac{1}{2}\left(\frac{\partial^2 \ln \psi(\bar{x}, \bar{y}, \bar{t})}{\partial \bar{x}^2} + \frac{\partial^2 \ln \psi(\bar{x}, \bar{y}, \bar{t})}{\partial \bar{y}^2}\right) \tag{2.25}$$

If the simple harmonic oscillator is on a time-invariant and uniform total potential, the periodic trajectory can be expected. Taking the wave function in the ground state $\psi_{00}$ as an example, it can be seen from Figure 2.1 that the trajectory is composed of multiple periodic interactions, rather than a general single periodic motion, and this phenomenon is caused by the continuous change of the total potential. In quantum Hamiltonian mechanics, because $\bar{x}, \bar{y}$ are complex, the originally time-invariant total potential is complexed, that is, $V_{Total}(\bar{x}_R, \bar{x}_I, \bar{y}_R, \bar{y}_I)$ expanded to be determined by four real variables. In other words, particles moving in a four-dimensional space can be projected onto a two-dimensional real number space. Usually, to analyze the movement of particles on the total potential of the real part, we must fix the imaginary part $\bar{x}_I, \bar{y}_I$ unchanged to see the distribution of the total potential from the real part $\bar{x}_R, \bar{y}_R$. But it is important to note that the total potential of the real part plane will change with the position of the imaginary part of the particle, which also represents the original time-varying potential in the complex space, just like the time-varying potential under the observation of the real number plane.

$$V_{Total}(\bar{x}_R, \bar{x}_I(\bar{t}), \bar{y}_R, \bar{y}_I(\bar{t})), \ -\infty \le \bar{x}_R, \bar{y}_R \le \infty \tag{2.26}$$

Therefore, the movement of particles in the complex space, due to the immeasurability of the dynamics of the imaginary part, will cause the total potential to change with time. This characteristic can explain why the trajectory exhibits a multi-periodic result.

From the observation of the wave function of the first excited state, take the same $\beta = 0.8, \gamma = 0.4$ as an example, and assume that the complex number space $\bar{x}_I = \bar{y}_I = 0$ is projected onto the real number plane under the condition of the total potential diagram, as shown in Figure 2.7.

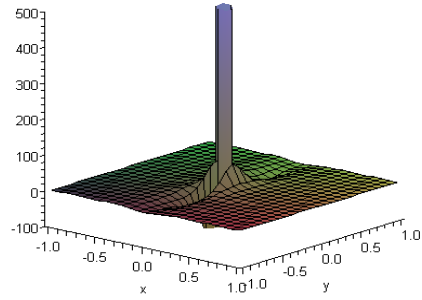

**Figure 2.7 When $\bar{x}_I = \bar{y}_I = 0$, the real part total potential map in the two-dimensional real number plane**

There is a large mountain columnar potential near the origin ($\bar{x}_R = 0$, $\bar{y}_R = 0$), which is caused by the quantum potential $Q(\bar{x}, \bar{y})$ in the middle area, while the external potential presents a smooth surface on the periphery. Moving particles to these areas with steep potentials, under the obvious strong change of the force, forced to transfer to other completely different directions due to slight disturbances in the current position, an irregular trajectory will arise accordingly. It forms the chaotic phenomenon we have observed; if the particles start from a region far away from the potential of the intermediate cliff, there will be no chaotic behavior.

## 3. Design and simulation analysis of quantum chaos controller

Now consider two identical nonlinear dynamic systems, one master system to control a slave system. The master system is also called the drive system, which can be defined as

$$\dot{\mathbf{x}} = \mathbf{A}\mathbf{x} + \mathbf{f}(\mathbf{x})$$
(3.1)

Where $\mathbf{x} = (x_1, x_2, \ldots, x_n)^T \in R^n$ is the state vector, and A is a linear coefficient $n \times n$ matrix, $\mathbf{f}$ is a nonlinear vector function. The slave system represents the response system and can also be defined as

$$\dot{\mathbf{y}} = \mathbf{A}\mathbf{y} + \mathbf{f}(\mathbf{y}) + \mathbf{u}(t)$$
(3.2)

$y = (y_1, y_2, \ldots, y_n)^T \in R^n$ is the state vector with the control input vector $\boldsymbol{u}(t) = (u_1(t), u_2(t), \ldots, u_n(t))^T \in R^n$. Then, we set that $\boldsymbol{e} = \boldsymbol{y} - \boldsymbol{x}$ is the difference between the state of the slave system and the state vector of the master system. Our goal of chaos synchronization is to design a controller $\boldsymbol{u}(t)$ so that the state vector of the slave system gradually approaches the state vector of the master system, limiting the tracking error $\boldsymbol{e}(t) = (e_1, e_2, \ldots, e_n)^T$ to zero: $\lim_{t \to \infty} \|\boldsymbol{e}(t)\| = 0$. The rate of variation of tracking error can be expressed as

$$\dot{\mathbf{e}} = \dot{\mathbf{y}} - \dot{\mathbf{x}} = \mathbf{A}\mathbf{e} + \mathbf{F}(\mathbf{x}, \mathbf{y}) + \mathbf{u}(t)$$
(3.3)

Where $F(x, y) = f(y) - f(x)$ is the difference of function in the nonlinear vectors of the two systems. According to the principle of active control, if we choose

$$\mathbf{u}(t) = \mathbf{H}(t) - \mathbf{F}(\mathbf{x}, \mathbf{y})$$
(3.4)

We can eliminate the nonlinear function in Eq. (3.3), which can be rewritten as

$$\dot{\mathbf{e}} = \mathbf{A}\mathbf{e} + \mathbf{H}(t)$$
(3.5)

If $\mathbf{H}(t)$ is regarded as a function of the tracking error $\mathbf{e}$, the dynamics added control can be viewed as a linear system with input $\mathbf{H}(t)$. If the feedback control of the closed-loop system can ensure stability, the tracking error can converge to zero. Then, we can construct $\mathbf{H}(t)$ from the sliding mode control:

$$\mathbf{H}(t) = \mathbf{K}w(t)$$
(3.6)

$\mathbf{K} = (k_1, k_2, \ldots, k_n)^T$ is the constant gain vector for adjusting the size of the state, and $w(t)$ is the input met the control gain switching condition, as in (1.1), it can be expressed as

$$w(t) = \begin{cases} w^-(t), & s(\mathbf{e}) > 0 \\ w^+(t), & s(\mathbf{e}) < 0 \end{cases}$$
(3.7)

$s(\boldsymbol{e})$ is the sliding surface defined in the space of the error state

$$s(\mathbf{e}) = c_1 e_1 + c_2 e_2 + \ldots + e_n = \mathbf{C}\mathbf{e}$$
(3.8)

$\mathbf{C} = [c_1, c_2, \ldots, c_{n-1}, 1]$ is a real vector, and in the vector $c_1, c_2, \ldots, c_{n-1}$ is exactly the coefficient $\lambda^{n-1} + c_{n-1}\lambda^{n-2} + \ldots + c_1$ in the Hurwitzian series. After combining Eq. (3.5) and Eq. (3.6), the tracking error expression can be obtained:

$$\dot{\mathbf{e}} = \mathbf{A}\mathbf{e} + \mathbf{K}w(t)$$
(3.9)

First, according to the balance control in the sliding mode, the condition $s(\boldsymbol{e}) = 0$ must be met when the state trajectory falls on, so $\dot{s}(\boldsymbol{e}) = 0$ from the relationship between Eq. (3.8) and Eq. (3.9), we can make

$$\dot{s}(\mathbf{e}) = \mathbf{C}[\mathbf{A}\mathbf{e} + \mathbf{K}w(t)] = 0$$
(3.10)

The above equation can be solved to obtain the balance control $w_{Eq}(t)$ as

$$w_{Eq}(t) = -(\mathbf{CK})^{-1}\mathbf{CAe}$$

(3.11)

With the condition that $(\mathbf{CK})^{-1}$ exist and $(\mathbf{A}, \mathbf{K})$ is controllable. Substituting Eq. (3.11) into Eq. (3.9), the state equation in sliding mode is

$$\dot{\mathbf{e}} = [\mathbf{I} - \mathbf{K}(\mathbf{CK})^{-1}\mathbf{C}]\mathbf{Ae}$$

(3.12)

Secondly, consider the design of switching control. We can design the switching control by the control law of normal number gain as

$$\dot{s} = -q \cdot \text{sign}(s) - rs$$

(3.13)

Eq. (3.13) uses an ideal switching function $sign(s)$, and the gain $q, r > 0$ is used to satisfy the proximation condition and sliding condition in the sliding mode to ensure that the sliding mode movement can occur. Finally, from Eq. (3.13), all inputs including balance control and switching control can be obtained as

$$w(t) = -(\mathbf{CK})^{-1}[\mathbf{C}(r\mathbf{I} + \mathbf{A})\mathbf{e} + q \cdot \text{sign}(s)]$$

(3.14)

In a real system, the control signal will be delayed due to the sampling time or the time of calculating the control loop. Therefore, it is impossible to switch the instantaneous time in practice, so that the trajectory of the system must bounce in the minimal space $\varepsilon$ on both sides $s > 0, s < 0$ of the sliding mode. causing high-frequency oscillations, which is called the chattering phenomenon. Many methods have been proposed to improve the chattering phenomenon, and we usually a saturation function $sat(s)$ to replace $sign(s)$:

$$w(t) = -(\mathbf{CK})^{-1}[\mathbf{C}(r\mathbf{I} + \mathbf{A})\mathbf{e} + q \cdot \text{sat}(s)]$$

(3.15)

The saturation function can be expressed by the following figure and Eq. (3.16)

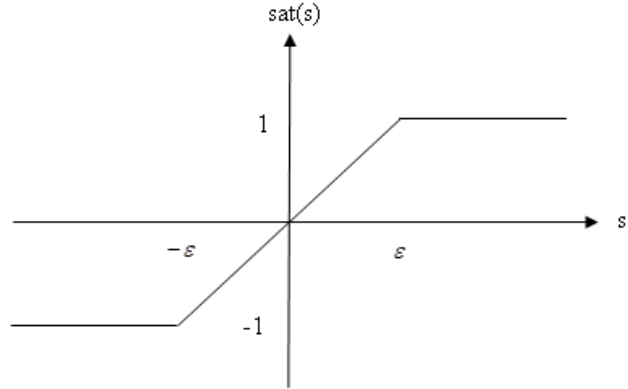

**Figure 3.1 sat($s$) function graph**

$$\text{sat}(s) = \begin{cases} \text{sign}(s), & |s| > \varepsilon \\ s \,/\, \varepsilon, & |s| \le \varepsilon \end{cases}$$

(3.16)

After selecting the control law, to verify the stability of the control system, we use the Lyapunov theorem to perform robust stability analysis. First, choose to have a Lyapunov function as follows:

$$\mathbf{V} = \frac{1}{2}s^2$$

(3.17)

The above equation ensures $V > 0$, and then take its derivative with respect to time

$$\dot{\mathbf{V}} = s\dot{s} = s\frac{\partial s}{\partial \mathbf{e}}\dot{\mathbf{e}} = s\mathbf{C}\dot{\mathbf{e}}$$

(3.18)

Substituting Eq. (3.9) (3.15) for Eq. (3.17), we can get

$$\dot{\mathbf{V}} = s\mathbf{C}[\mathbf{Ae} + \mathbf{K}w(t)] = s\mathbf{C}\left\{\mathbf{Ae} - \mathbf{K}(\mathbf{CK})^{-1}[\mathbf{C}(r\mathbf{I} + \mathbf{A})\mathbf{e} + q \cdot \text{sat}(s)]\right\}$$

$$= -rs^2 - sq \cdot \text{sat}(s)$$

(3.19)

When $e \ne 0$, it can always maintain a negative value and satisfy $\dot{V} = s\dot{s} < 0$. Therefore, the designed sliding mode controller can indeed stabilize the closed-loop system.

Let's take the model mentioned in section 2 as an example, and use the control mentioned above to achieve synchronization. Rewrite Eq. (2.22) to the following:

$$\begin{bmatrix} \dfrac{d\bar{x}}{dt} \\ \dfrac{d\bar{y}}{dt} \end{bmatrix} = \begin{bmatrix} -2ia_3 & a_4 \\ a_4 & -2ia_5 \end{bmatrix}\begin{bmatrix} \bar{x} \\ \bar{y} \end{bmatrix} + \begin{bmatrix} \dfrac{-ia_1}{a_1\bar{x} + ib_1\bar{y}} \\ \dfrac{b_1}{a_1\bar{x} + ib_1\bar{y}} \end{bmatrix}$$

(3.20)

From complex mechanics, the two complex numbers $\bar{x}, \bar{y}$ are transformed into $\bar{x} = \bar{x}_R + i\bar{x}_I$, $\bar{y} = \bar{y}_R + i\bar{y}_I$ and substituting to Eq. (3.20), which can be extended to the four-dimensional state vector of the real and imaginary parts of the two axes. To distinguish the master system and the slave system, we make the state vector $\bar{x}_R = x_{1R}, \bar{x}_I = x_{1I}, \bar{y}_R = x_{2R}, \bar{y}_I = x_{2I}$ in the master system; the corresponding state vector of the slave system is $\bar{x}_R = y_{1R}, \bar{x}_I = y_{1I}, \bar{y}_R = y_{2R}, \bar{y}_I = y_{2I}$. Then, we can obtain the complete differential equations of the master system and the slave system, Eq. (3.21) (3.22):

$$\dot{\mathbf{x}} = \begin{bmatrix} \dfrac{dx_{1R}}{dt} \\ \dfrac{dx_{1I}}{dt} \\ \dfrac{dx_{2R}}{dt} \\ \dfrac{dx_{2I}}{dt} \end{bmatrix} = \begin{bmatrix} 0 & 2a_3 & a_4 & 0 \\ -2a_3 & 0 & 0 & a_4 \\ a_4 & 0 & 0 & 2a_5 \\ 0 & a_4 & -2a_5 & 0 \end{bmatrix} \begin{bmatrix} x_{1R} \\ x_{1I} \\ x_{2R} \\ x_{2I} \end{bmatrix} + \mathbf{f}(x_{1R}, x_{1I}, x_{2R}, x_{2I})$$

(3.21)

The right side of the equal sign is a $4 \times 4$ linear matrix $\boldsymbol{A}$ and a nonlinear function $\boldsymbol{f}(x_{1R}, x_{1I}, x_{2R}, x_{2I})$. The non-linear function part can be completely expressed as

$$\begin{bmatrix} \dfrac{-(a_1^2 x_{1I} + a_1 b_1 x_{2R})}{a_1^2(x_{1R}^2 + x_{1I}^2) + b_1^2(x_{2R}^2 + x_{2I}^2) + 2a_1 b_1(x_{1I}x_{2R} - x_{1R}x_{2I})} \\ \dfrac{-(a_1^2 x_{1R} - a_1 b_1 x_{2I})}{a_1^2(x_{1R}^2 + x_{1I}^2) + b_1^2(x_{2R}^2 + x_{2I}^2) + 2a_1 b_1(x_{1I}x_{2R} - x_{1R}x_{2I})} \\ \dfrac{-b_1^2 x_{2I} + a_1 b_1 x_{1R}}{a_1^2(x_{1R}^2 + x_{1I}^2) + b_1^2(x_{2R}^2 + x_{2I}^2) + 2a_1 b_1(x_{1I}x_{2R} - x_{1R}x_{2I})} \\ \dfrac{-(b_1^2 x_{2R} + a_1 b_1 x_{1I})}{a_1^2(x_{1R}^2 + x_{1I}^2) + b_1^2(x_{2R}^2 + x_{2I}^2) + 2a_1 b_1(x_{1I}x_{2R} - x_{1R}x_{2I})} \end{bmatrix}$$

Similarly, the slave system can be additionally described by the following equation:

$$\dot{\mathbf{y}} = \begin{bmatrix} \dfrac{dy_{1R}}{dt} \\ \dfrac{dy_{1I}}{dt} \\ \dfrac{dy_{2R}}{dt} \\ \dfrac{dy_{2I}}{dt} \end{bmatrix} = \begin{bmatrix} 0 & 2a_3 & a_4 & 0 \\ -2a_3 & 0 & 0 & a_4 \\ a_4 & 0 & 0 & 2a_5 \\ 0 & a_4 & -2a_5 & 0 \end{bmatrix} \begin{bmatrix} y_{1R} \\ y_{1I} \\ y_{2R} \\ y_{2I} \end{bmatrix} + \mathbf{f}(y_{1R}, y_{1I}, y_{2R}, y_{2I}) + \mathbf{u}(t)$$

(3.22)

According to the chaotic synchronization principle, the tracking error vector $\boldsymbol{e}$ is composed of the slave system minus the four states of the master system, and the controller $\boldsymbol{u}(t) = \boldsymbol{K}w(t) - \boldsymbol{F}(\boldsymbol{x}, \boldsymbol{y})$ can be determined by Eq. (3.4), and the function $\boldsymbol{F}(\boldsymbol{x}, \boldsymbol{y})$ is $\boldsymbol{f}(y_{1R}, y_{1I}, y_{2R}, y_{2I}) - \boldsymbol{f}(x_{1R}, x_{1I}, x_{2R}, x_{2I})$. Also, the tracking error change rate between the two systems can be obtained from Eq. (3.3)

$$\dot{\mathbf{e}} = \begin{bmatrix} \dfrac{de_{1R}}{dt} \\ \dfrac{de_{1I}}{dt} \\ \dfrac{de_{2R}}{dt} \\ \dfrac{de_{2I}}{dt} \end{bmatrix} = \begin{bmatrix} 0 & 2a_3 & a_4 & 0 \\ -2a_3 & 0 & 0 & a_4 \\ a_4 & 0 & 0 & 2a_5 \\ 0 & a_4 & -2a_5 & 0 \end{bmatrix} \begin{bmatrix} e_{1R} \\ e_{1I} \\ e_{2R} \\ e_{2I} \end{bmatrix} + \mathbf{F}(\mathbf{x}, \mathbf{y}) + \mathbf{u}(t)$$

(3.23)

However, for adjusting the $4 \times 1$ input weight vector $\boldsymbol{K}$, the second and fourth parameters should be set to zero, mainly because we cannot observe the dynamics of the projection on the imaginary part, so we cannot exert control outside. Then there is the design problem of sliding surface $s(\boldsymbol{e}) = c_1 e_1 + c_2 e_2 + c_3 e_3 + e_4 = \boldsymbol{C}\boldsymbol{e}$ in sliding mode. The real coefficients $[c_1, c_2, c_3]$ in the row vector To satisfy the asymptotic stability condition of the Hurwitz criterion, we can judge that the stable condition is from the Routh Table, and the final control signal $w(t)$ is set as in equation (3.15).

Suppose that the goal is to make a quantum system with chaotic phenomena (slack system) track the desired stable period under the same system (master system), under the action of the controller, the tracking error can be returned to zero, and the original chaotic trajectory of the particle can be restored To periodicity, the controlled closed-loop system will also be stable, which is also the core of chaos control in this research. Next, the control simulation is divided into two parts: The first part is the test of different starting positions, discussing the response of the disturbance in the complex space to the control trajectory, and unifying the fixed sliding surface $\boldsymbol{C} = [1,2,2,1]$, weight vector $\boldsymbol{K} = [1,0,1,0]^T$, and controller gain parameters $q = 1, r = 1$. The second part is about the design of the controller parameters, by adjusting the gain value to judge the quality of the controller.

In the following simulation, we discuss the chaotic trajectory caused by different starting positions and control the particles to enter a stable periodic trajectory according to the difference of the chaotic strength. Assuming that the starting positions $[x_R, x_I, y_R, y_I] = [0.4,0,0.4,0], [1.1,0,0,0], [1.1,0,1,0]$ are given to the secondary system in sequence, the result is shown in the figure below, and you can see the disorderly trajectory:

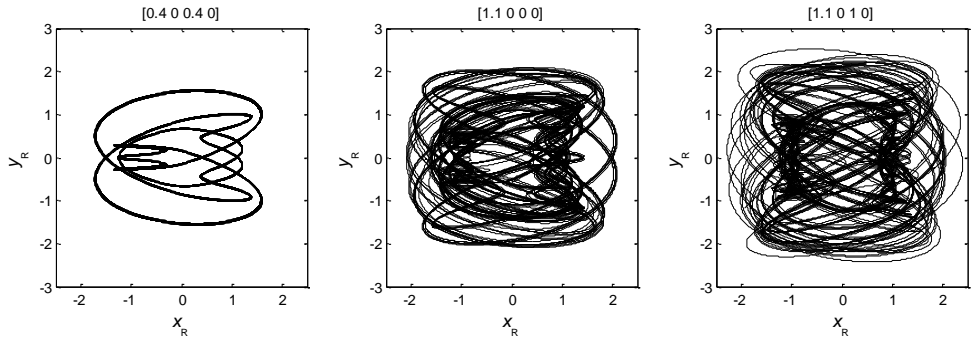

**Figure 3.2 Chaotic trajectory under different initial conditions**

The master system needs to generate an external signal to drive the slave system so that the particles tend to a stable periodic trajectory. We take $[x_R, x_I, y_R, y_I] = [2,0,2,0]$ in the master system as the starting position of the state, we want to track. In the simulation results after the controller is applied, Figure 3.3 and Figure 3.4 respectively show the dynamic response of particles in the $x$ and $y$ direction. The original three sets of different chaotic trajectories gathered to the periodic trajectory we specified in about 10 seconds, completing the chaos synchronization control. Besides, the tracking error response in Figure 3.5 and Figure 3.6 shows that the error will converge to 0 after a certain time. These results show that chaos control has achieved the expected goal.

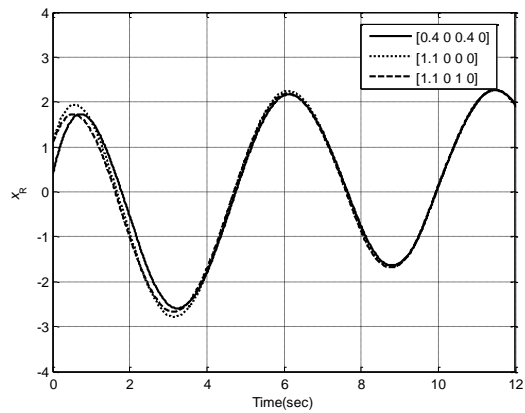

**Figure 3.3 Chaos synchronization response of three sets of chaotic trajectories on the x-axis**

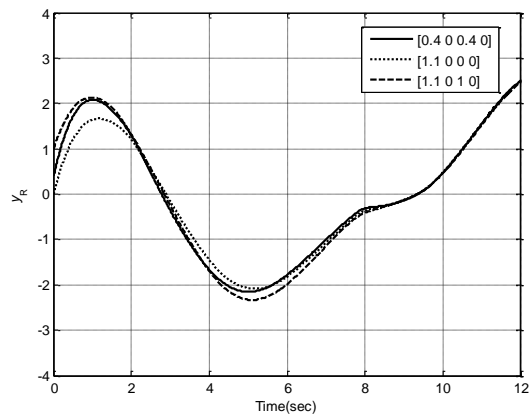

**Figure 3.4 Chaos synchronization response of three sets of chaotic trajectories on the y-axis**

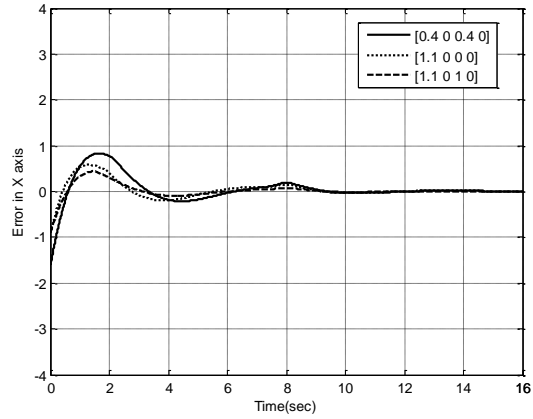

**Figure 3.5 Error response of three groups of chaotic trajectories on the x-axis**

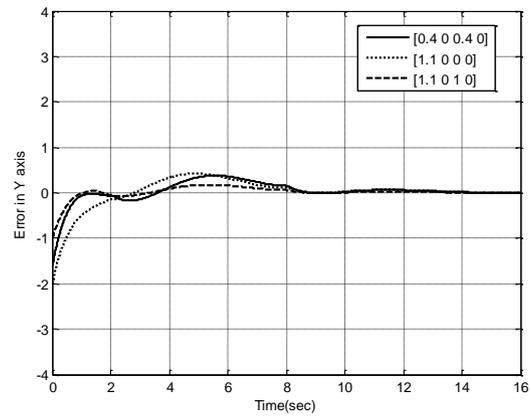

**Figure 3.6 Error response of three groups of chaotic trajectories on the y-axis**

From the three groups of trajectories, we selected the group [1.1,0,1,0] with the strongest chaos phenomenon and observed the plane trajectory and input response after control. The result is shown in Figure 3.7. The difference from the irregular trajectory in Figure 3.2, the controlled particles tend to be stable and move regularly, and the control input also presents a smooth curve. Since the saturation function in the controller effectively reduces the chattering phenomenon.

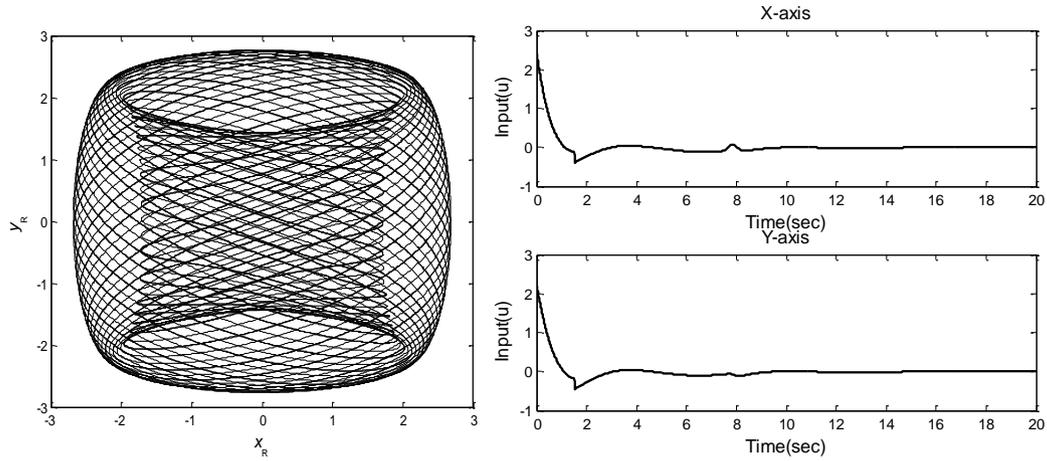

**Figure 3.7 Control input response and two-dimensional plane trajectory**

Figure 3.8 shows the state of the control system moving on the phase plane. The phase plane trajectory curve falls on the sliding surface $s(e) = 0$ within a short response of 1 to 2 seconds, which meets the basic requirements of sliding mode control.

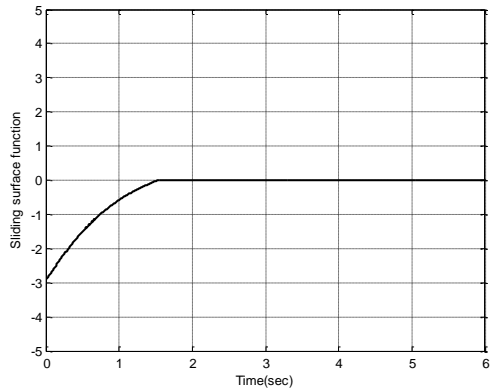

**Figure 3.8 Sliding plane function response**

We can also control the same chaotic trajectory to different target trajectories. Considering the chaotic trajectory generated by the starting point [1.1,0,1,0], we want to control it to the periodic target trajectory generated by the three initial points [0,0,0.4,0],[3,0,0,0],[1.5,0,0.8,0] to test its synchronization ability. Figure 3.7 and Figure 3.8 show the dynamic response in the corresponding direction. The solid line is the controlled response (slave system) dynamic, and the dashed line is the target dynamic of the request (master system). Depending on the trajectory selection, the time to reach the synchronization state varies slightly from long to short. Then Figure 3.9 and Figure 3.10 show the tracking error response in each direction. We can see that the error curve will eventually converge to 0 through time.

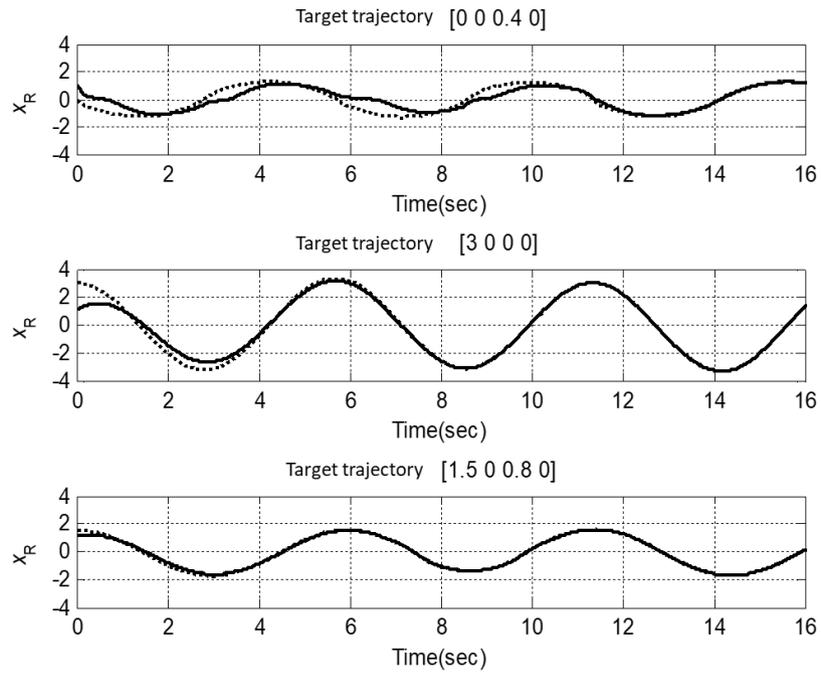

**Figure 3.9 Dynamic response on the x-axis for three control periods**

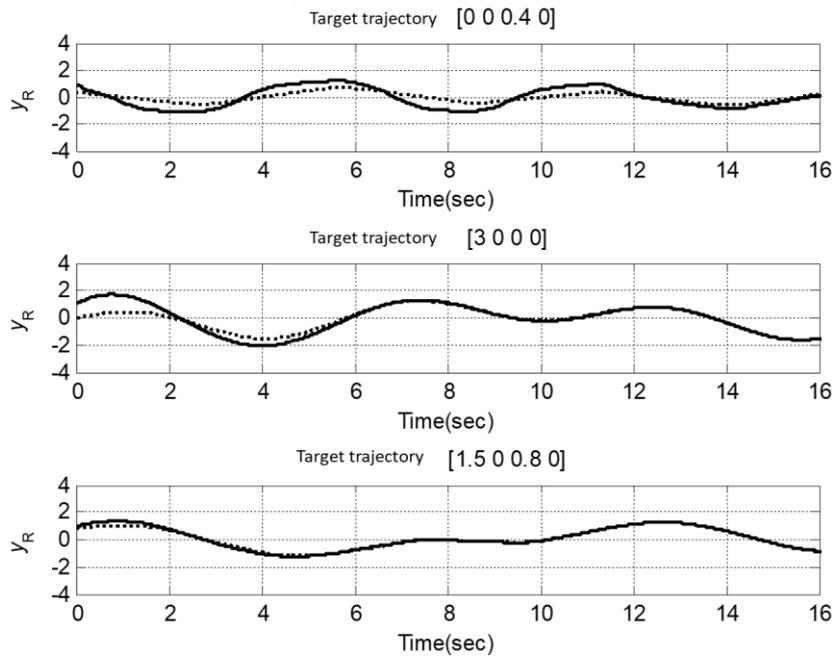

**Figure 3.10 Dynamic response on the y-axis for three control periods**

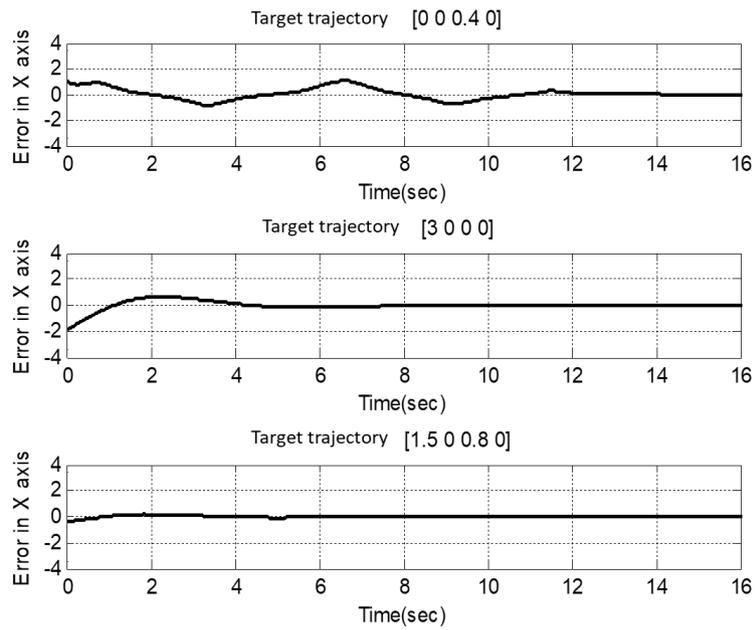

**Figure 3.11 Error response on the x-axis for three control periods**

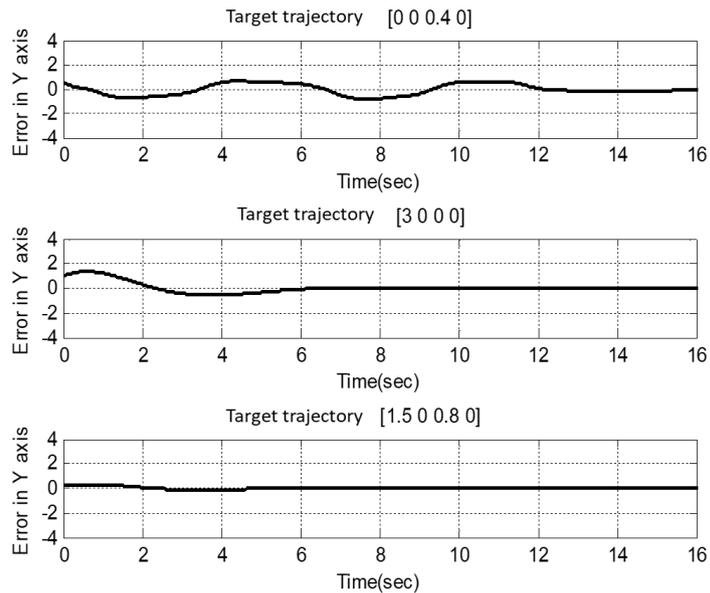

**Figure 3.12 Error response on the y-axis for three control periods**

Figure 3.13 shows the three types of periodic target trajectories obtained after chaos control. Regardless of the target trajectory, after the particles are controlled, they can move to a stable periodic trajectory and make regular motions, which is different from the chaotic trajectory before control.

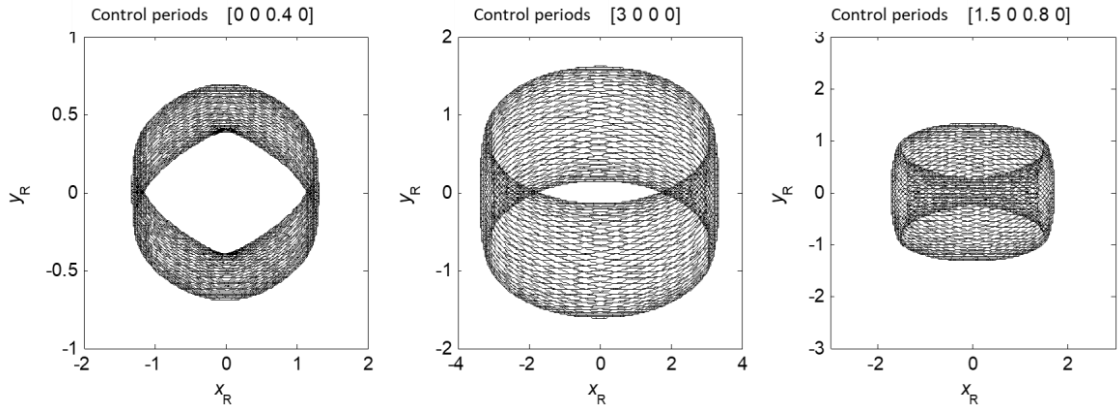

**Figure 3.13 Respectively control to three types of periodic trajectories (target trajectories)**

The above-mentioned trajectories are all caused by the disturbance of the initial position of the real part. Here we further discuss the trajectory caused by the disturbance of the imaginary part. Following the discussion of the strong chaotic phenomenon in Section 2, the phenomenon of trajectory divergence caused by the unobservable change of the starting point of the imaginary part has been revealed in Figure 2.2. Applying the same control method to strong chaos control, the dynamic response in Figure 3.14 starts from the same initial point, and the trajectory is synchronized after 12 seconds. Besides, Figure 3.15 shows that the tracking error can also converge to 0 to ensure the convergence of the state. Let's observe the shape of the trajectory after input response and control in Figure 3.16. Compared with Figure 3.3-3.12, it is obvious that the control input fluctuates greatly and the curve is less smooth. For the plane trajectory, it can smoothly converge to periodicity, which is clearly distinguished in the simulation in Figure 3.16.

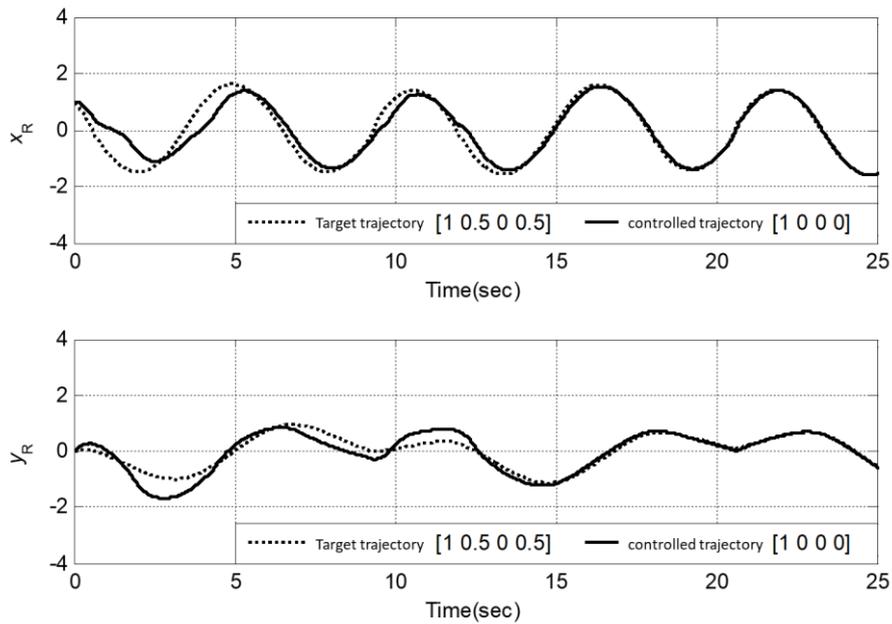

**Figure 3.14 Strong chaotic dynamic response**

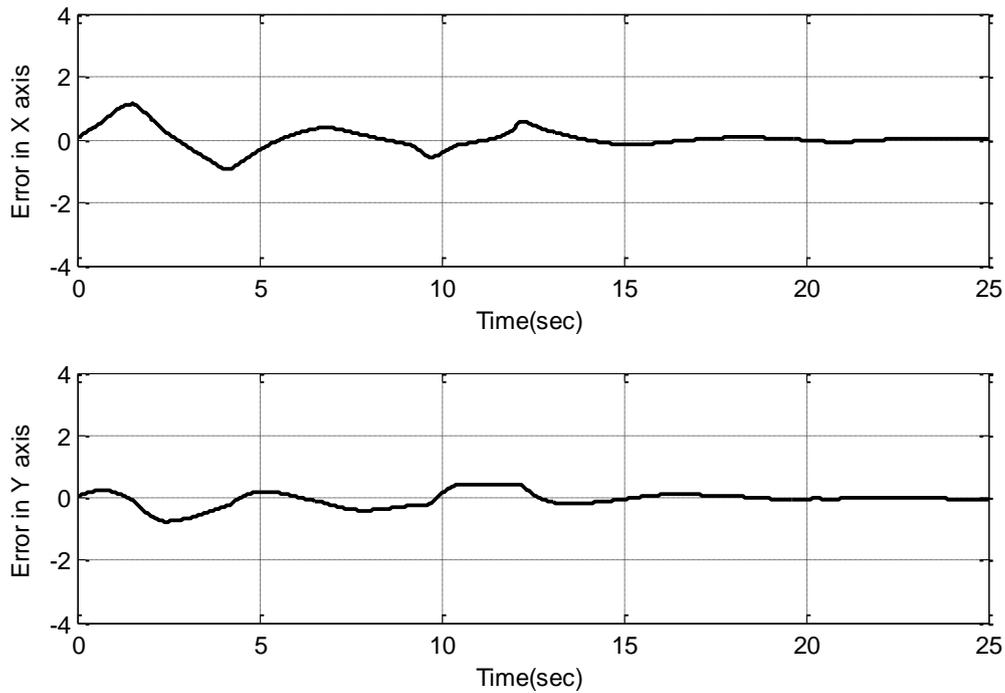

**Figure 3.15 Strong chaotic error response**

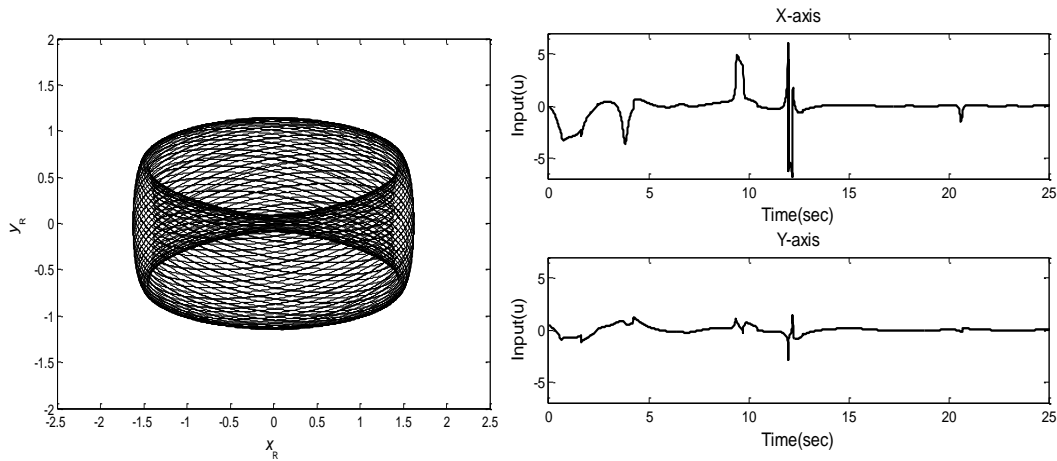

**Figure 3.16 Strong chaotic control input response and two-dimensional plane trajectory**

## 4. The influence of controller parameter adjustment

When designing the controller, to ensure the establishment of the Lyapunov stability. To demonstrate the unique robustness of the control system, the gain parameters $q$ and $r$ must be greater than zero, simulations were performed under fixed parameters before. The following is to compare the benefits of quantum control from the simulation results by adjusting the size of the parameters.

First of all, we fix the value of $r$, and make the value of 1, 3, 6 change sequentially, and at the same time take a chaotic trajectory control as an example. The following simulation results show that the trajectory starting from the initial position [1.1,0,1,0] is synchronized to the trajectory starting from [2,0,2,0]. In the error response of Figure 4.1 and Figure 4.2, as the gain parameter $q$ increases, the convergence speed of the tracking error becomes slightly faster.

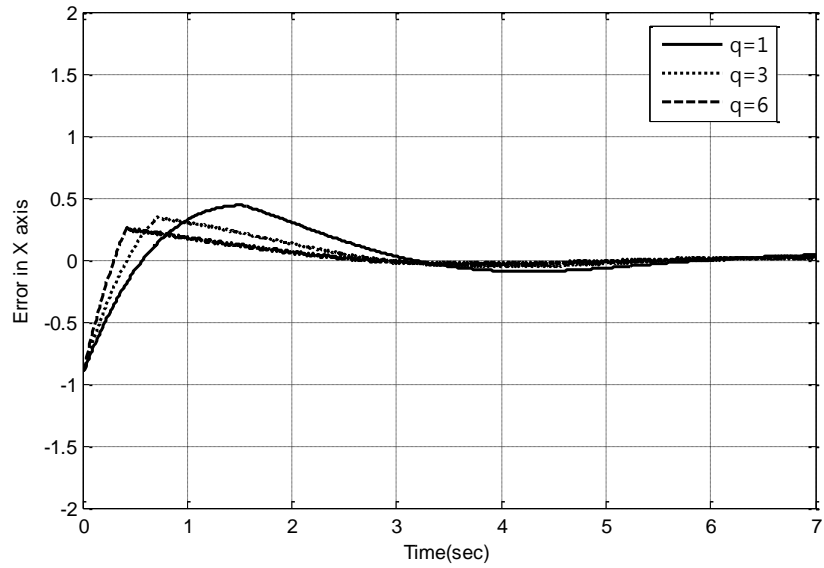

**Figure 4.1 The influence of parameter *q* changes on the x-axis tracking error**

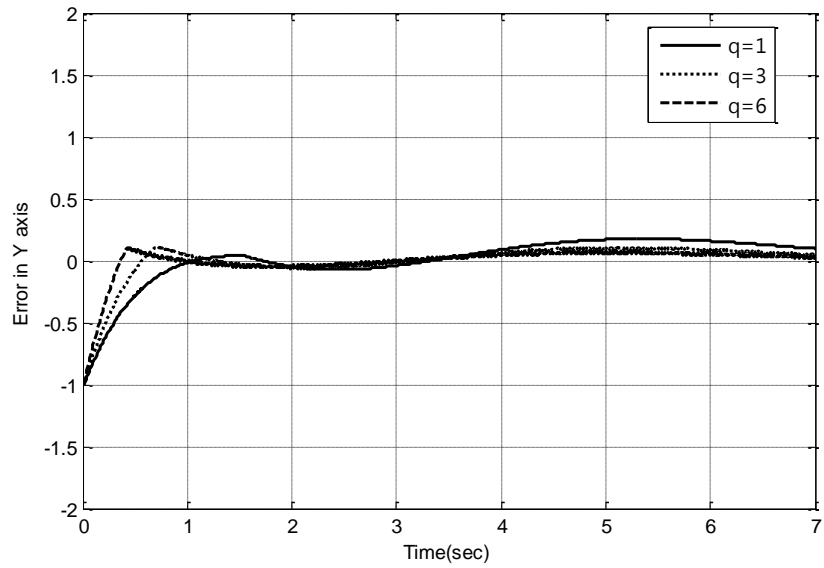

**Figure 4.2 The influence of parameter *q* changes on the y-axis tracking error**

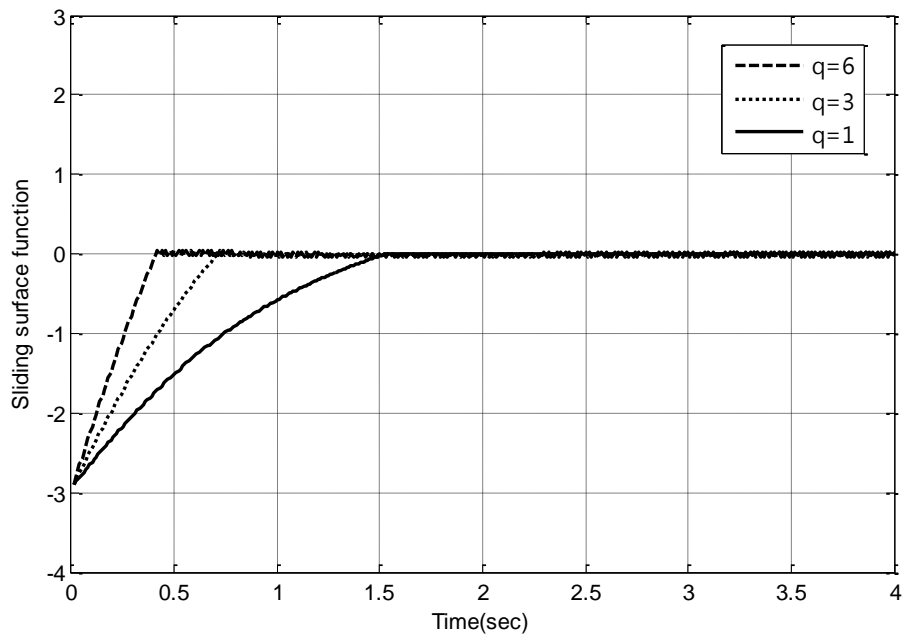

**Figure 4.3 The influence of parameter *q* changes on sliding surface function**

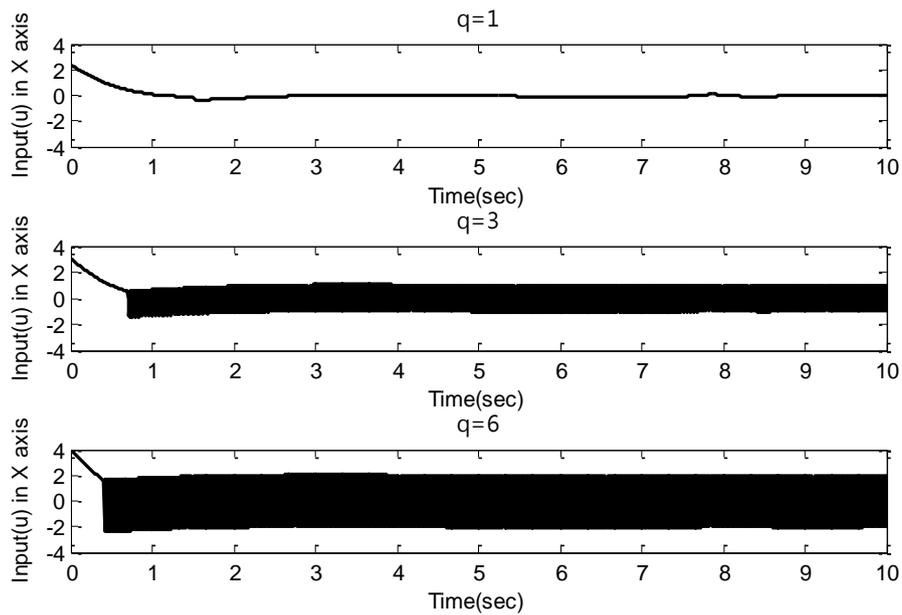

**Figure 4.4 The effect of parameter *q* changes on x-axis input**

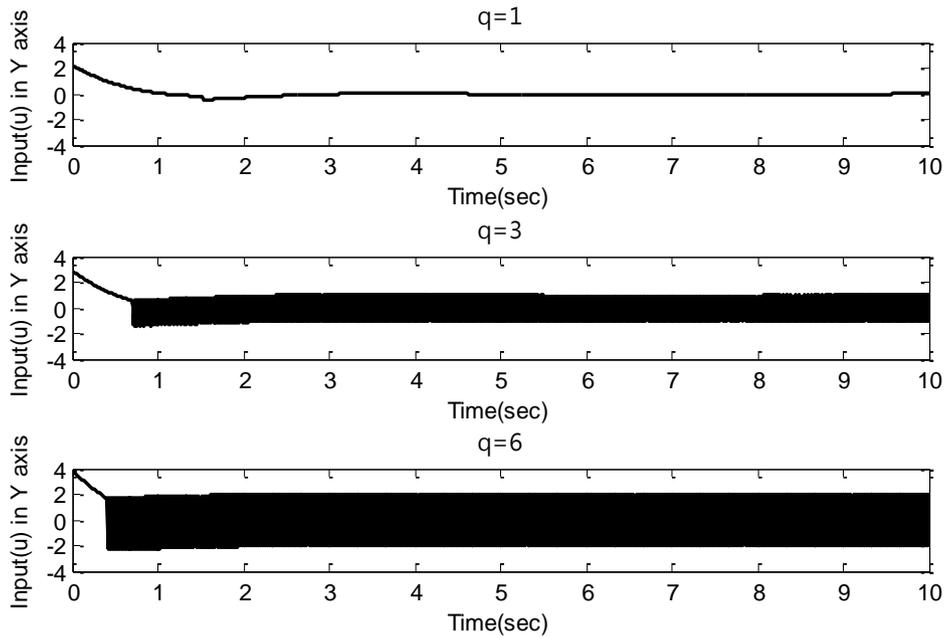

**Figure 4.5 The effect of parameter *q* changes on y-axis input**

Figure 4.3 shows the response of the sliding function with parameters $q$. When the larger $q$ is, the trajectory approaches the sliding surface faster, and the curve processes in an approximate straight line. In Figure 4.4 and Figure 4.5, it can be seen that the control input has become larger and exhibits high-amplitude changes. This result is not unexpected, because the parameter $q$ is mainly to adjust the gain of the switching control, the larger the value, the faster the speed of the state to the sliding surface, causing the chattering phenomenon. Therefore, although the increase $q$ can make the error convergence speed faster, it will also cause excessive energy consumption.

Then, let's discuss the change of value $r$. Fixing parameters $q$, we use the same changes as 1, 3, and 6. The result of the simulation is similar to that of changing the parameters. In Figure 4.6 and Figure 4.7, the convergence speed of the tracking error tends to become faster in the 2-second response section.

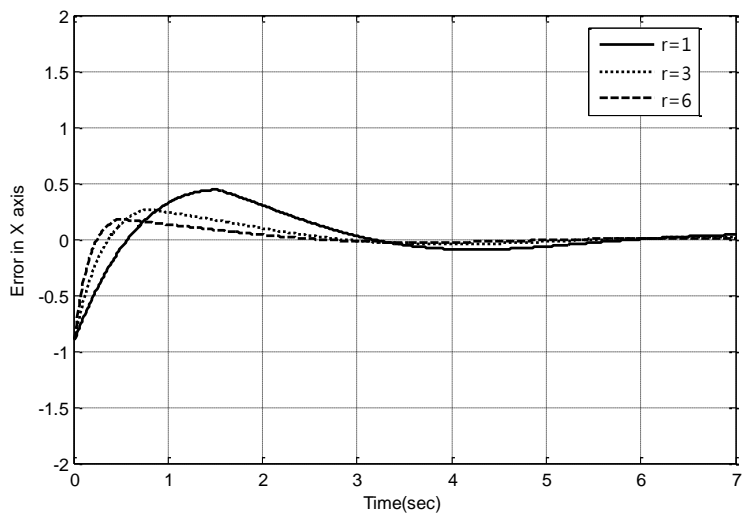

**Figure 4.6 The influence of parameter *r* changes on the x-axis tracking error**

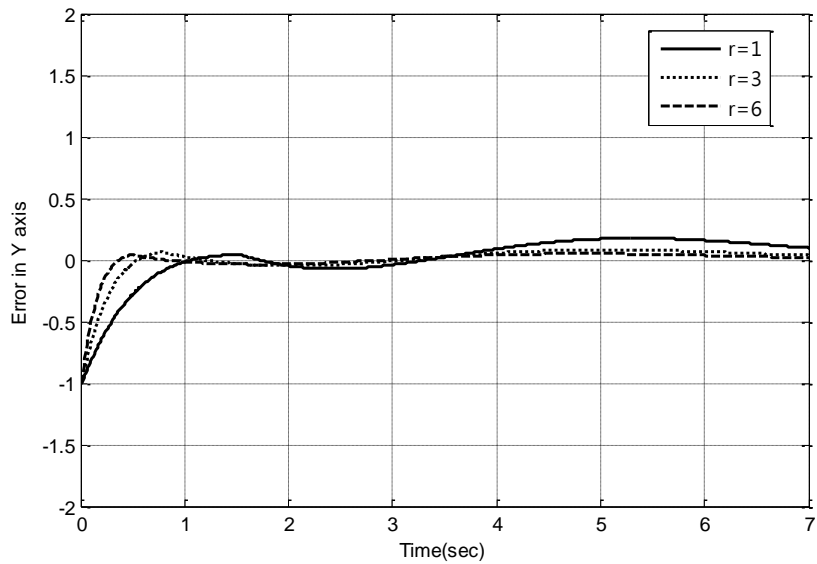

**Figure 4.7 The influence of parameter $r$ changes on the y-axis tracking error**

Figure 4.8 shows that in the short-term response of the sliding function, as the parameter increases, it approaches the sliding surface earlier. This is compared with Figure 4.3 of the parameter $q$. The difference is that there is no chattering phenomenon, and the curve falls smoothly into the sliding surface. Surface. Looking at the input size in Figure 4.9 and Figure 4.10, the input tends to be fastly stable in 5 seconds of response. Comparing Figure 4.4 and Figure 4.5 respectively, there is no high-amplitude oscillation during adjustment $q$. This shows that adjustment $r$ can avoid excessive energy consumption. Overall, adjusting the ratio of the $r$ to $q$ in the controller parameters can achieve better results. In addition to accelerating the convergence speed of the response, it can also reduce the chattering phenomenon in the sliding mode and save energy.

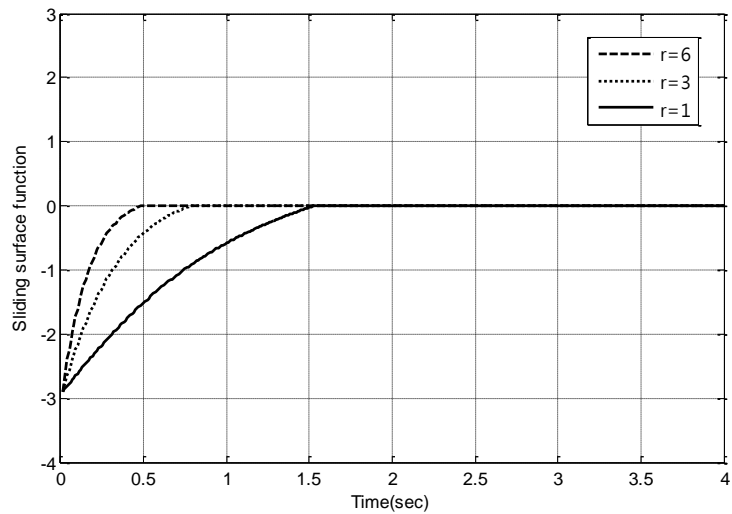

**Figure 4.8 The influence of parameter $r$ changes on sliding surface function**

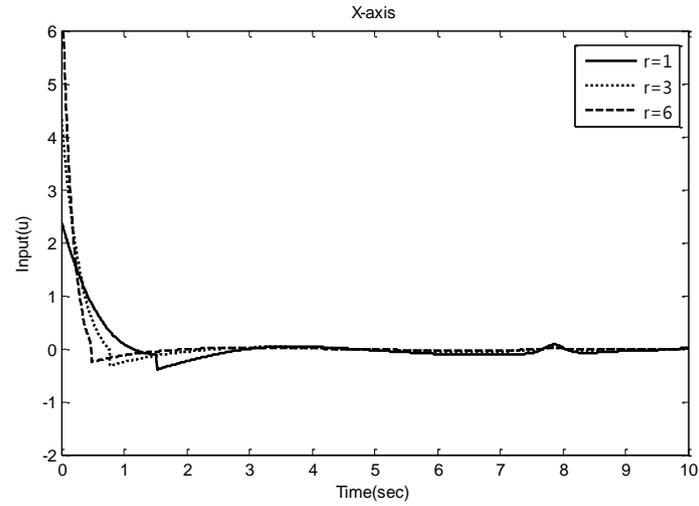

**Figure 4.9 The effect of parameter $r$ changes on x-axis input**

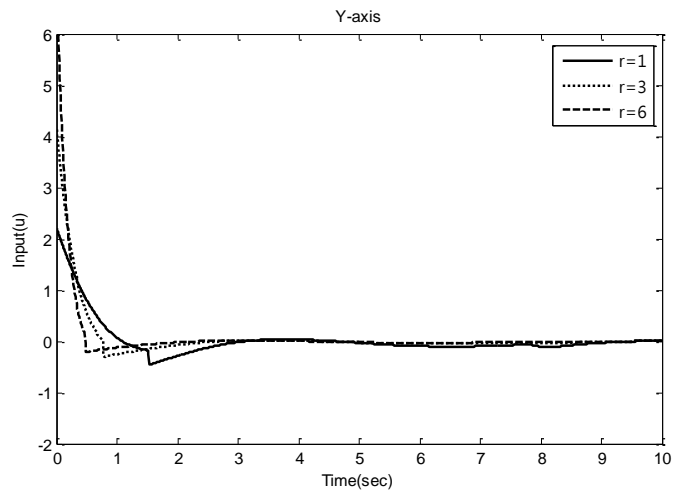

**Figure 4.10 The effect of parameter $r$ changes on y-axis input**

## 5. Verification and discussion

In this section, we apply the chaos analysis method to verify the effectiveness of chaos control. Start with the spectrum analysis and observe the spectrum of three different chaotic trajectories. From these figures, we can understand that the spectrum intensity shows different distributions due to the different strengths of the chaos. When the chaotic phenomenon is more significant, the power intensity distribution is more broaden and more disorderly. As shown in Figure 5.1-5.3.

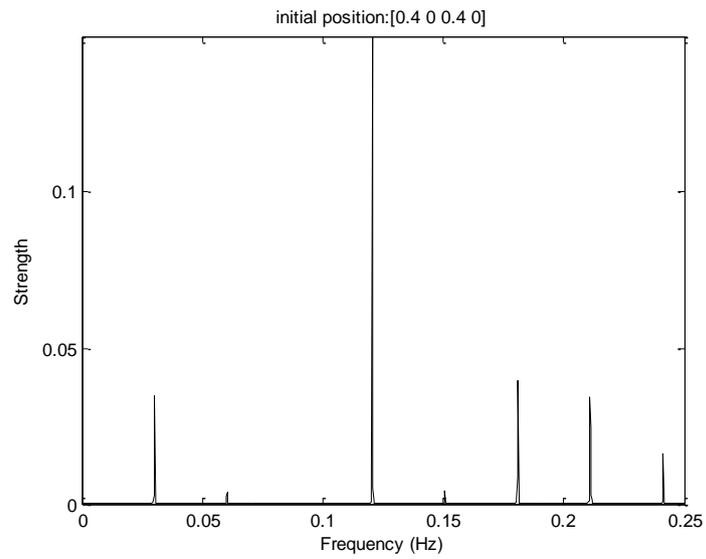

**Figure 5.1 Spectrum analysis of the trajectory starting from the initial position [0.4 0 0.4 0] before being controlled**

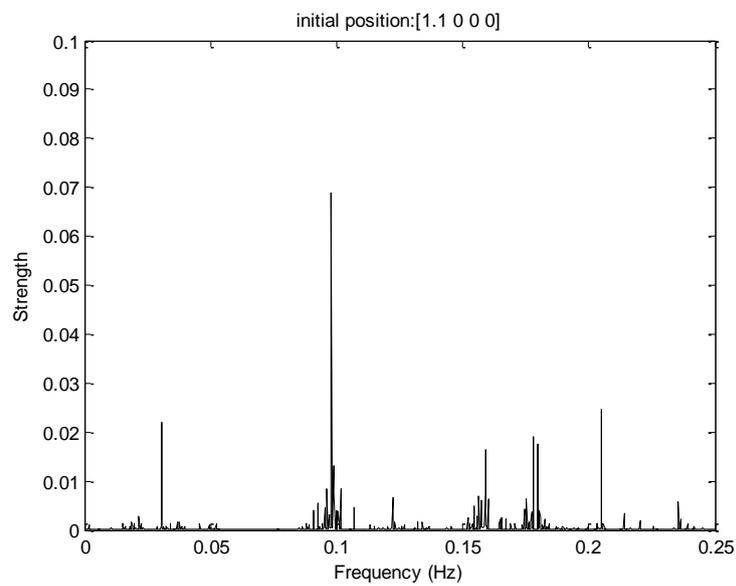

**Figure 5.2 Spectrum analysis of the track starting from the initial position [1.1 0 0 0] before being controlled**

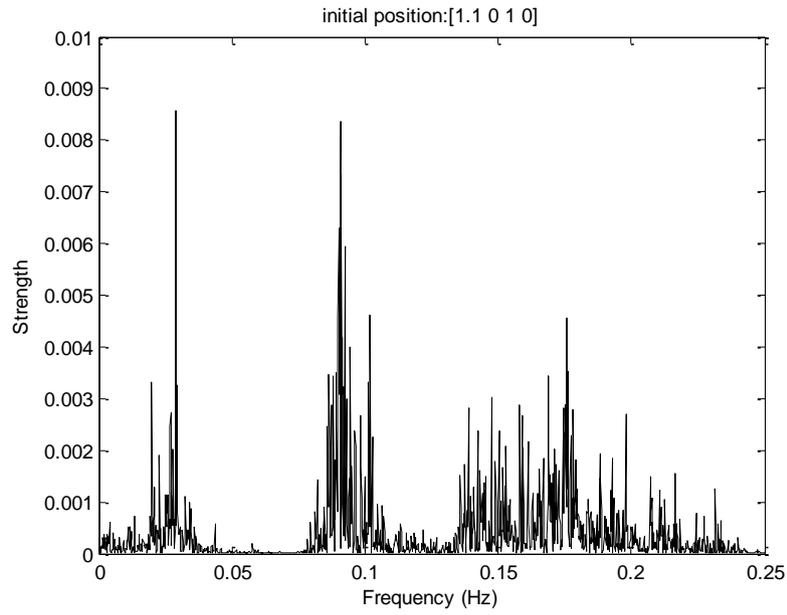

**Figure 5.3 Spectrum analysis of the trajectory starting from the initial position [1.1 0 1 0] before being controlled**

If one expects to quantify the difference in the degree of chaos in different chaotic systems, the Lyapunov exponent is the most convincing index. In this article, to ensure that sufficient information can be retrieved from the system, we uniformly select the reconstruction dimension $d = 6$, adjust the time delay window parameter range from 5 to 15, and calculate with the time series of the real axis to obtain uncontrolled chaos measurement at each initial position. From Figure 5.4-5.6, they depict the Lyapunov exponent under different delay time windows in the reconstruction phase space. Each curve in the figure represents the change value under different dimensions, the higher the curve represents the higher the embedded dimension, and also the sequential dimension is $d = 1 \sim 6$.

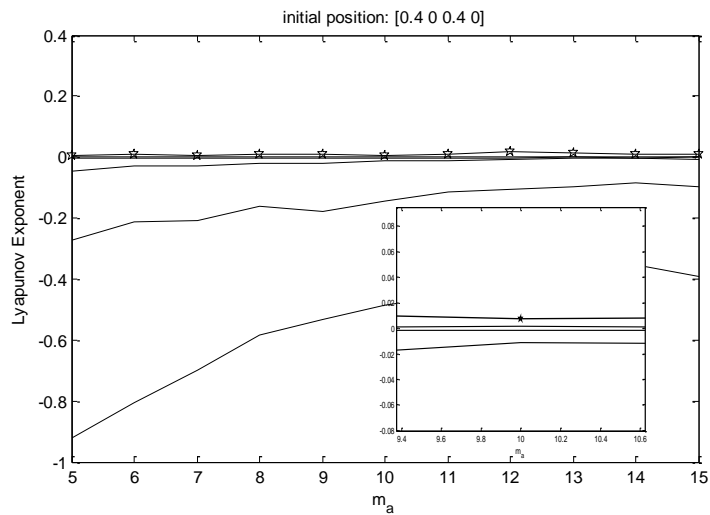

**Figure 5.4 Lyapunov exponent = 0.01 at the initial position [0.4 0 0.4 0] before being controlled**

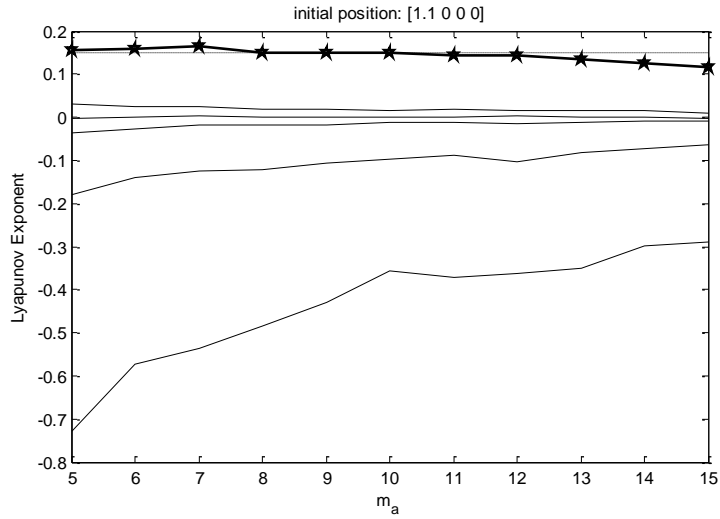

**Figure 5.5 Lyapunov exponent = 0.148 at the initial position [1.1 0 0 0] before being controlled**

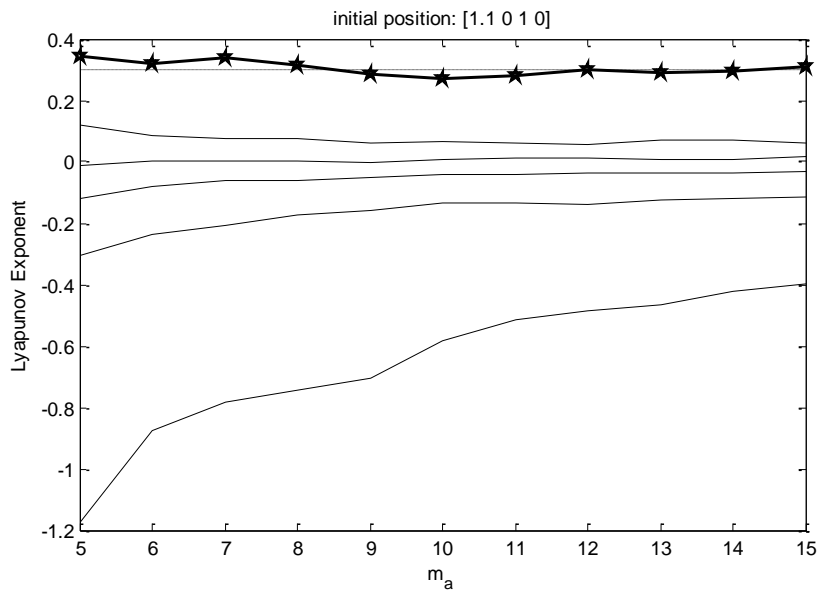

**Figure 5.6 Lyapunov exponent = 0.294 at the initial position [1.1 0 1 0] before being controlled**

Taking a closer look at Figure 5.4, we can see that the quantified Lyapunov exponent converges to 0.01, showing weak chaos. Then, the Lyapunov exponent in Figure 5.5 has a higher value 0.148, showing that the dynamic chaotic effect is significant. Lyapunov exponent value 0.294 in Figure 5.6 can be judged that the chaotic effect of this system is more significant. Therefore, the trajectories starting from different starting positions have different chaotic dynamics.

After the controlled quantum trajectory, we also use spectrum analysis and Lyapunov exponent to verify whether the system is controlled. The target trajectory of a given chaos control is still set as the trajectory starting from the initial position [2,0,2,0]. The spectrograms of the first three different chaotic trajectories after control are shown from Figure 5.7-5.9. The spectrograms of the controlled system are almost the same, and they all show peaks at two specific vibration frequencies. The irrational value obtained by dividing the two peak frequencies shows that it is a quasi-periodical trajectory, which is very different from the original chaotic trajectory.

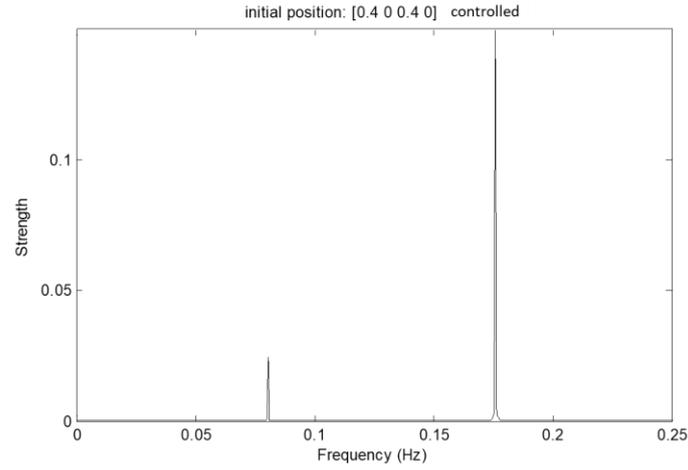

**Figure 5.7 Spectrum analysis of the controlled track starting from the initial position [0.4 0 0.4 0]**

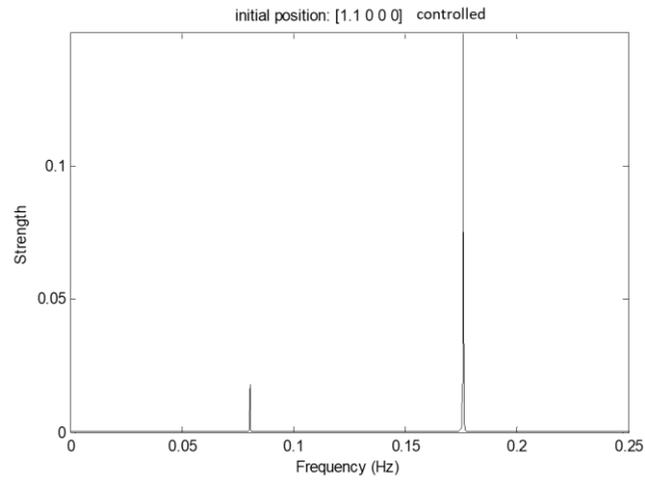

**Figure 5.8 Spectrum analysis of the controlled track starting from the initial position [1.1 0 0 0]**

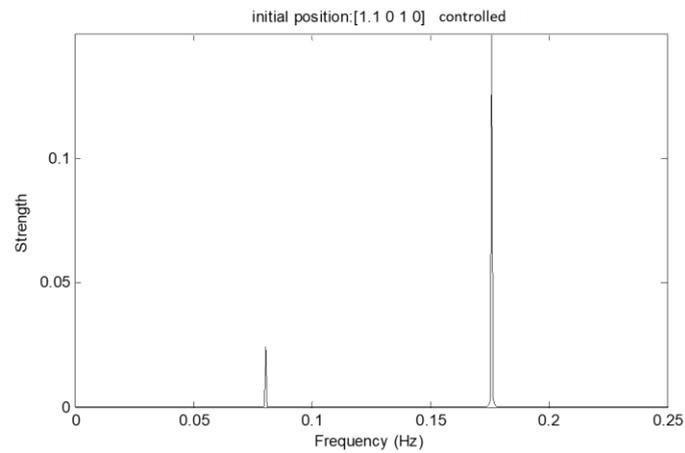

**Figure 5.9 Spectrum analysis of the controlled track starting from the initial position [1.1 0 1 0]**

Similarly, check Lyapunov exponent. Figure 5.10 shows the change of Lyapunov exponent after the chaotic system is controlled.

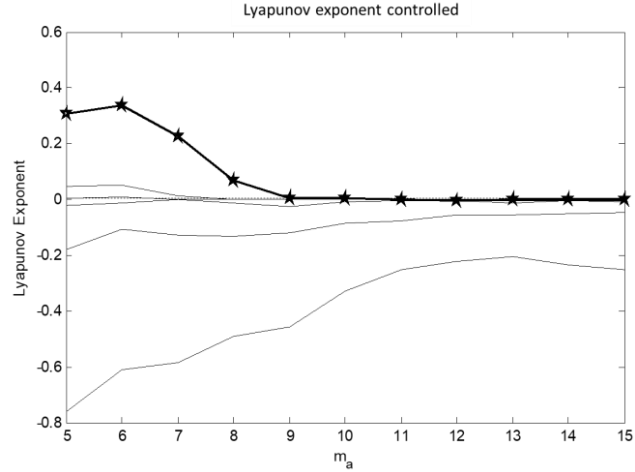

**Figure 5.10 Change of Lyapunov exponent after each chaotic trajectory is controlled**

Comparing Lyapunov exponent in Figure 5.4-5.6, the value read in Figure 5.10 is smaller and approaching zero. This means that the trajectory after control is no longer diverge randomly. This quantitative index provides favorable evidence for the suppression of chaos and also means that the quantum chaos control we designed is indeed feasible. The following table lists the Lyapunov exponent before and after chaos control compiled in section 4.

**Table 5.1 Lyapunov exponent before and after 5.2.1 control for Figure 3.3 to 3.7**

| Initial position (slave system) | uncontrolled | controlled | Initial position (master system) |
|---|---|---|---|
| [0.4, 0, 0.4, 0] | 0.01 | 0.0015 | |
| [1.1, 0, 0, 0] | 0.148 | 0.004 | [2, 0, 2, 0] |
| [1.1, 0, 1, 0] | 0.294 | 0.005 | |

**Table 5.2 Lyapunov exponent and after 5.2.2 control for Figure 3.9 to 3.13**

| Initial position (slave system) | uncontrolled | controlled | Initial position (master system) |
|---|---|---|---|
| | | 0.015 | [0, 0, 0.4, 0] |
| [1.1, 0, 1, 0] | 0.294 | 0.008 | [3, 0, 0, 0] |
| | | 0.0028 | [1.5, 0, 0.8, 0] |

**Table 5.3 Lyapunov exponent before and after control for Figure 3.14 to 3.16**

| Initial position (slave system) | Uncontrolled | controlled | Initial position (master system) |
|---|---|---|---|
| [1, 0, 0, 0] | 0.2735 | 0.008 | [1, 0.5, 0, 0.5] |

## 6. Conclusion

This research mainly discusses quantum chaos. Taking the two-dimensional charged anisotropic simple harmonic oscillator in a uniform magnetic field as an example. It uses complex mechanics to model the quantum system, and then projects the dynamic variables from the complex space to real numbers space. The important phenomena observed can be summarized as follows:

(1) From the chaotic trajectory Figure 2.2, it can indeed be seen that the particles exhibit irregular chaotic motion, and this interpretation of the dynamic trajectory of quantum mechanics can be analyzed using the classical concept of chaos, and the existence of quantum chaos phenomenon can be confirmed by the traditional chaotic analysis method.

(2) The strength of quantum chaos also depends on the initial position of the particle. If the starting position is close to a place where the qubit potential fluctuates sharply, the chaotic phenomenon becomes stronger.

(3) We can set the disturbance of the initial position of the imaginary part in the complex space to explain the origin of the strong chaotic phenomenon.

In the quantum chaos control part, the sliding mode controller is used to control the quantum chaos based on the quantum system model constructed above. After the simulation analysis results, the following key points are integrated:

(1) Under the fixed controller parameters, sequentially test the chaotic trajectories starting from different initial positions in the complex number space. As a result, the original chaotic motion can be turned into a periodic motion, which can be seen from the figure in Sections 3 with its strong effect.

(2) In Section 4, through the convergence speed of the tracking error, and the size of the control input, the influence of the controller parameters on the effectiveness of chaos control is evaluated.

(3) At the end of this article, the section 5, we analyze the change of the frequency band before and after the control and the decrease of Lyapunov exponent, confirming that the quantum chaos control proposed in this paper is indeed feasible.

REFERENCES

[1] G. M. Huang, T. J. Tarn, 1983, "On the controllability of quantum mechanical systems" Journal of Mathematics and Physics, 24:2603-2618.

[2] R.S. Judson, 1992, "Teaching lasers to control molecules", Phys. Rev. Lett. 68,1500.

[3] A. Assion, T. Baumert, M. Bergt, T. Brixner, B. Kielfer, V. Seyfried, M. Strehle, and G. Gerber, 1998, "Control of chemical reactions by feedback-optimized phase-shaped femtosecond laser pulses", Science 282,919.

[4] J. Kunde, B. Baumann, S. Arlt, 2001, June, "Optimization of adaptive feedback control for ultrafast semiconductor spectroscopy", Optical Society of America, Vol.18, No.6,

[5] A.C. Doherty, K. Jacobs, 1999, "Feedback-control of quantum systems using continuous state-estimation", Phys. Rev. A,60, 2700.

[6] M. Yanagisawa, H. Kimura, 2003, Dec, "Transfer function approach to quantum control part I: Dynamics of Quantum Feedback System", IEEE Transactions on Automatic Control, vol. 48.

[7] M. Yanagisawa, H. Kimura, 2003, Dec, "Transfer function approach to quantum control part II: Control Concepts and Applications", IEEE Transactions on Automatic Control, vol. 48.

[8] Shi-Ming Huang, 2005, "Robust Control of molecular motion," Department of Aeronautics and Astronautics, National Cheng Kung University

[9] L.M. Pecora, T.M. Carroll, 1990, "Synchronization in chaotic systems " Phys. Rev. Lett. 64:821-824.

[10] M.T. Yassen, 2005, "Controlling chaos and synchronization for new chaotic system using linear feedback control."Chaos, Solitons & Fractals 26:913-920.

[11] S. Chen, J. L, 2002, "Synchronization of an uncertain unified system via adaptive control."Chaos, Solitons & Fractals 14:643-647.

[12] H.N. Agiza, M.T. Yassen, 2001, "Synchronization of Rossler and Chen chaotic dynamical systems using active control ."Phys. Lett. A 278 191-197.

[13] Tavazoei, Mohammad Saleh & Haeri, Mohammad, 2008. "Synchronization of chaotic fractional-order systems via active sliding mode controller," Physica A: Statistical Mechanics and its Applications,, vol. 387(1), pages 57-70.

[14]S. Boccaletti, J. Kurths, G. Osipov, D. Valladares and C. Zhou, "The synchronization of chaotic systems", Physics Reports, vol. 366, no. 1-2, pp. 1-101, 2002.

[15] C. Yang, Z. Ge, C. Chang and S. Li, "Chaos synchronization and chaos control of quantum-CNN chaotic system by variable structure control and impulse control", Nonlinear Analysis: Real World Applications, vol. 11, no. 3, pp. 1977-1985, 2010.

[16] S. Kallush and R. Kosloff, "Mutual influence of locality and chaotic dynamics on quantum controllability", Physical Review A, vol. 86, no. 1, 2012.

[17] W. Li, C. Li and H. Song, "Criterion of quantum synchronization and controllable quantum synchronization based on an optomechanical system", Journal of Physics B: Atomic, Molecular and Optical Physics, vol. 48, no. 3, p. 035503, 2015.

[18] D. Dong and I. Petersen, "Sliding mode control of quantum systems", New Journal of Physics, vol. 11, no. 10, p. 105033, 2009.

[19] D. Dong and I. Petersen, "Sliding mode control of two-level quantum systems", Automatica, vol. 48, no. 5, pp. 725-735, 2012.

[20] D. Dong and I. Petersen, "Notes on sliding mode control of two-level quantum systems", Automatica, vol. 48, no. 12, pp. 3089-3097, 2012.

[21]S. Chegini and M. Yarahmadi, "Quantum sliding mode control via error sliding surface", Journal of Vibration and Control, vol. 24, no. 22, pp. 5345-5352, 2018.

[22] C. Edwards, S. Spurgeon, 1998 "Sliding mode Control: Theory and Applications", London: Taylor and Francis.

[23] Yang, C.D., 2006, "Quantum Hamilton mechanics: Hamilton equations of Quantum Motion, Origin of Quantum Operators, and Proof of Quantization Axiom", Annals of Physics. Vol. 321, pp. 2876-2926.

[24] Yang, C.D., 2006, "Modeling Quantum Harmonic Oscillator in Complex Domain", Chaos, Solitons and Fractals, Vol. 30, pp.342-362.

[25] Yang, C.D., 2007, " Quantum Motion in Complex Space", Chaos, Solitons and Fractals, Vol. 33, pp.1073-1092.

[26] Yang, C.D., 2008, "Trajectory Interpretation of the Uncertainty Principle in 1D Systems Using Complex Bohmian Mechanics", Physics Letters A, Vol. 372, pp.6240-6253.

[27] Yang, C.D., 2010, " Complex Mechanics", Asian Academic Publisher, ISSN 2077-8139, Hong Kong.

[28] H. Goldstein, 1980, "Classical Mechanics", Chapter 10, 2nd Ed., Addison- Wesley Publishing Company.

[29] O. Dippel, P. Schmelcher, and L.S. Cederbaum, 1994, "Charged Anisotropic Harmonic Oscillator and The Hydrogen Atom in Crossed Fields", Physics Review A, Vol.49, pp.4415-4429.

[30] Chia-Hung Wei, 2009, "A Study on Quantum Chaos and Quantum Probability from the Viewpoint of Complex Quantum Trajectories," Department of Aeronautics and Astronautics, National Cheng Kung University

[31] G. Arfken, 1985 "Hypergeometric Functions", §13.5 in Mathematical Methods for Physicists, 3rd Ed. Academic Press, pp.748-752.